\documentclass[journal]{IEEEtran}
\usepackage{color}
\usepackage{amsmath}
\usepackage{amssymb}
\usepackage{graphicx}
\usepackage{bm}
\usepackage{cases}
\newtheorem{lemma}{Lemma}
\newtheorem{theorem}{Theorem}
\ifCLASSINFOpdf

\fi

\begin{document}

\title{Uplink Enhancement for Low-Altitude UAVs via Active RISs Under Existing Cellular Infrastructure}

\author{Weicong~Chen,~\IEEEmembership{Member,~IEEE,} Chao-Kai~Wen,~\IEEEmembership{Fellow,~IEEE}, Wankai~Tang,~\IEEEmembership{Member,~IEEE,} Yiping~Zuo,~\IEEEmembership{Member,~IEEE,}   Xiao~Li,~\IEEEmembership{Member,~IEEE,}  
        and~Shi~Jin,~\IEEEmembership{Fellow,~IEEE}% <-this % stops a space
\thanks{{Weicong Chen, Wankai Tang, Xiao Li, and Shi Jin are with the National Mobile Communications Research Laboratory, Southeast University,
		Nanjing, 210096, P. R. China (e-mail: cwc@seu.edu.cn; tangwk@seu.edu.cn; li\_xiao@seu.edu.cn; jinshi@seu.edu.cn).} }
	\thanks{Chao-Kai Wen is with the Institute of Communications Engineering, National Sun Yat-sen University, Kaohsiung 80424, Taiwan. (e-mail: chaokai.wen@mail.nsysu.edu.tw). }
	\thanks{Yiping Zuo is with the school of Computer Science, Nanjing University of Posts and Telecommunications, Nanjing, China (e-mail: zuoyiping@njupt.edu.cn). }
	}

\markboth{Journal of \LaTeX\ Class Files,~Vol.~14, No.~8, August~2015}%
{Shell \MakeLowercase{\textit{et al.}}: Bare Demo of IEEEtran.cls for IEEE Journals}

\maketitle

\begin{abstract}
The proliferation of low-altitude aerial Internet of Things applications, such as UAV-based sensing/monitoring, demands high-performance uplinks, yet terrestrial-optimized cellular networks with downtilted base station (BS) antennas form coverage holes that severely degrade aerial uplink. To address this challenge without modifying infrastructure, active reconfigurable intelligent surfaces (RISs) are introduced. Unlike existing studies with complicated active RIS optimizations, we conduct a detailed noise analysis to simplify the transmission design. By deploying one active RIS per BS sector to fully exploit multi-sector reception, we develop a joint sector maximal signal combining scheme that improves signal quality through diversity gain. Active RISs introduce amplifier noise, raising questions about their net benefit. We derive an upper bound on the power ratio between RIS-introduced noise and BS-inherent noise, revealing that the impact of RIS noise is negligible under typical operating conditions. Based on this insight, we maximize a derived lower bound of the received signal-to-noise ratio by sequentially designing the amplitude and phase of reflection coefficients with closed-form expressions. Simulation results validate the theoretical analysis on negligible RIS-introduced noise in typical systems and demonstrate that under the same array gain, a larger RIS size rather than a higher reflection gain is preferred for enhancing the low-altitude uplink transmission.

\end{abstract}


\IEEEpeerreviewmaketitle

\section{Introduction}

\IEEEPARstart{T}{he} past decades have {seen major progress} in wireless communications, aiming to achieve near-ubiquitous coverage for terrestrial users through {carefully} planned cellular networks \cite{ubiquitous} \cite{ubiquitous-2}. However, the emerging aerial Internet of Everything (IoE) paradigm \cite{AIoT} \cite{IoE}, empowered by uncrewed aerial vehicles (UAVs) and electric vertical take-off/landing (eVTOL), demands a vertical extension of wireless coverage into low-altitude airspace. Unlike traditional terrestrial communications characterized by high downlink and low uplink traffic, low-altitude IoT scenarios demand a reversed data transmission pattern. For instance, UAVs performing tasks such as agricultural monitoring \cite{UAV-agri}, power line inspection \cite{UAV-PLI}, and urban flood area detection \cite{UAV-floods} often require real-time uplink transmission of high-definition (HD) video streams while receiving minimal downlink control signals. This asymmetric traffic profile, however, is poorly supported by existing networks, where base station (BS) antenna arrays are typically configured with downtilt angles to optimize terrestrial coverage \cite{downtilt}. Such configurations result in coverage holes in the low-altitude airspace, severely limiting the capacity for high-rate uplink from aerial devices.

With the growing demand for low-altitude wireless coverage \cite{LAE-1,LAE-2,LAE-3}, researchers have been exploring innovative solutions to fill the coverage gap in this challenging environment. The authors in \cite{Cover-JLi23} utilized a flexible and scalable cell-free radio access network to support low-altitude three-dimensional (3D) communications, achieving a trade-off between coverage performance and system scalability. To alleviate challenges in detecting and tracking low-flying targets due to aerial blind areas, a frequency-diversity subaperturing multiple-input multiple-output (MIMO) radar was proposed in \cite{Cover-XLi18}, where the low-altitude beam coverage performance was optimized by exploiting the additional degrees of freedom in the joint spatial and spectral domains. Considering that using the existing coverage structure would result in problems of under-coverage and over-coverage, the authors in \cite{Cover-JLu24} proposed a triangular prism-based 3D coverage structure and a ground-to-air coverage expansion method, ensuring seamless 3D coverage by minimizing the spatial overlap of different beams. Utilizing legacy terrestrial BSs, a coordinated multi-point transmission based on the geometry of Poisson-Delaunay triangulation was developed in \cite{Cover-YLi23}, providing seamless connectivity for UAVs. {Recently, fluid/movable antenna systems \cite{KK-FAS}\cite{RZhang-MA} have been proposed, enabling dynamic control of antenna tilt to achieve full-space coverage for both terrestrial and low-altitude airspace \cite{LAE-MA}\cite{LAE-FAS}. While deploying new BSs for aerial coverage or adjusting BS antennas/beams to an upward tilt could theoretically address the coverage issue in low-altitude airspace, this approach requires a trade-off between the performance of aerial users and terrestrial users and disrupts existing network planning by introducing varying inter-cell interference.}

Thus, cost-effective and interference-aware alternatives are preferred. Among emerging technologies, the reconfigurable intelligent surface (RIS) is a promising enabler for dynamic control of electromagnetic (EM) propagation environments \cite{SRE-1,SRE-2}. Passive RISs have been widely applied in various wireless scenarios, including cell-free networks \cite{P-cellfree}, over-the-air computation \cite{P-Aircomp}, channel estimation \cite{P-CE,cus_LFA}, device identification \cite{P-secu}, {integrated sensing and communication \cite{P-ISAC}, and coverage enhancement \cite{5G-JLi22}--\cite{CoverRIS-PSA24}. The RIS-enabled coverage enhancement in 5G commercial networks operating in sub-6 GHz was carried out in \cite{5G-JLi22} and \cite{5G-JSanS24}, where significant system improvements were observed in various urban scenarios. In existing millimeter-wave systems, where coverage issues are more pronounced, RISs have demonstrated even more substantial performance gains in enhancing coverage \cite{5G-HChoi24, 5G-NOk25}. The real-world test results in \cite{5G-HChoi24} showed signal enhancement of up to 20 dB in a weak signal area. In addition to existing networks, RISs were deployed to assist aerial BSs in eliminating coverage blind spots on uneven terrain \cite{CoverRIS-AVS23}. Utilizing the mobility and flexibility of UAVs, the authors in \cite{CoverRIS-PSA24,R2-1,R2-2} proposed mounting the RIS on the UAV, enabling coverage in areas that are difficult to reach with traditional wireless systems. Beyond terrestrial coverage enhancement, passive RISs have also been exploited for aerial interference mitigation \cite{R1-1,R1-2}. Positioning RISs to face upwards in multi-BS systems, inter-cell interference of cellular-connected UAVs can be significantly suppressed by optimizing coordinated beamforming and phase shift \cite{R1-2}. All the aforementioned studies focused on passive RIS-assisted systems, confronting substantial challenges posed by multiplicative path loss \cite{WTang1}.}

 In these contexts, active RISs are favored because they not only reflect/refract signals but also amplify them \cite{ARIS-LWu} \cite{ARIS-refract}, thereby mitigating the impact of path loss. The authors in \cite{ARIS-DM24} developed the first active RIS-assisted MIMO orthogonal frequency-division multiplexing (OFDM) prototype compliant with the 5G New Radio standard, demonstrating that a smaller active RIS has greater potential compared to a larger passive RIS. {For secure communications}, as detailed in \cite{ARIS-DWang24}, an aerial active RIS was introduced to maximize the secrecy rate. When the active RIS is introduced to integrate sensing and communication systems, the radar signal-to-interference ratio (SNR) was maximized in \cite{ARIS-ZYu24} by jointly optimizing the beamforming matrix at a dual-function radar-communication BS and the reflection vectors at the RIS. {Considering that signal amplification also incurs energy consumption, joint transmit beamforming and energy-included RIS reflection matrix design algorithm was proposed in \cite{ARIS-YMa23} to maximize energy efficiency.} Given that perfect channel state information (CSI) is unattainable due to estimation and feedback errors \cite{CSI-CE,CSI-LF,CSI-error},  the authors in \cite{ARIS-JYang25} proposed a robust transmission design to maximize the minimum achievable rate of each user under a statistical CSI error model. The aforementioned studies, ranging from prototype implementation to theoretical analysis, demonstrated the application potential of active RISs in wireless systems. By deploying active RISs within existing networks, stable virtual line-of-sight (LoS) links can be established between aerial devices and BSs, effectively filling the coverage hole in the low-altitude domain without modifying existing network settings. To the best of the authors' knowledge, no existing work has yet explored how to use active RISs to enhance low-altitude uplink transmission in the presence of down-tilted BS antennas.

In this study, we introduce active RISs to enhance uplink transmission in low-altitude economies with a tri-sector BS. {Unlike passive RISs-assisted cellular-connected UAV scenarios \cite{R1-1}\cite{R1-2}, active RISs amplify the uplink signal but simultaneously introduce amplifier noise, which complicates the SNR analysis at the tri-sector BS.} To gain insight into the uplink SNR for the UAV, we derive upper bounds for the power ratio between the RIS-introduced noise and the BS-inherent noise, demonstrating that the RIS amplifier noise can be neglected in typical scenarios. Based on this finding, we achieve the optimal RIS reflection coefficient design to maximize the lower bound of the uplink received SNR. The key contributions of this research are as follows:

\begin{itemize}
	\item \emph{A joint sector maximal signal combining (JS-MSC) for uplink enhancement.} Unlike terrestrial communications, the uplink signal from UAVs in low-altitude communications can be received by all antenna sectors of the BS through the reflection of distributed RISs. To enhance uplink transmission, we propose a JS-MSC reception scheme at the BS, utilizing the multi-sector diversity gain of the BS. Based on this reception scheme, closed-form expressions of the average signal power and noise power are derived to guide the reflection design of active RISs.
	\item {\emph{An upper bound on noise power ratio for noise comparison.}} Both active and passive RISs can enhance signal quality, but active RISs introduce additional amplifier noise that depends on the specific reflection design. {Existing studies \cite{ARIS-LWu}--\cite{ARIS-YMa23} and \cite{ARIS-JYang25}--\cite{ARIS-power-2} have not clarified how the RIS-introduced noise and BS-inherent noise jointly affect the received SNR. To address this research gap, we derive an upper bound for the power ratio of these two types of noise.} Although this upper bound is loose, theoretical results show that it remains small under typical network parameters. This finding suggests that, in practical scenarios, the noise introduced by active RISs with moderate amplification gain can be neglected.
	\item \emph{An optimal reflection design for maximizing the lower bound of received SNR.} Based on the analysis of the noise power ratio, a lower bound for the SNR of the JS-MSC receiver is derived. To maximize the SNR with the reflection design of active RISs, we decouple the phase and amplitude of the reflection coefficients and optimize them using {a sequential optimization approach}. The derived results indicate that, due to large-scale fading, the power of the signal reaching the RIS is relatively low, and thus employing the maximum reflection amplitude does not exceed the power constraint of the active RIS.
	\item \emph{Comprehensive simulations for validation of the proposed uplink enhancement scheme.} Numerical results verify the impact of various parameters, such as the maximum reflection amplitude, RIS size, and path loss, on the noise power ratio as expressed in the derived upper bound. Under optimal reflection configurations, the SNR gain achieved by active RISs is significantly greater than that of passive RISs. Moreover, in the considered scenario, a larger RIS size rather than a higher reflection gain is preferred for enhancing uplink transmission.
\end{itemize}

The remainder of this paper is structured as follows: Section \ref{sec:2} introduces the basic system model of this research. Section \ref{sec:3} analyzes the relationship between the RIS-introduced noise and the BS-inherent noise. Section \ref{sec:4} maximizes the lower bound of SNR {by sequentially optimizing} the phase and amplitude of the reflection coefficients. Numerical results for the enhanced uplink transmission are provided in Section \ref{sec:5}. Finally, we conclude this study in Section \ref{sec:6}. 

%\emph{Notations}: Lowercase and uppercase letters denote vectors and matrices, respectively. The transpose and conjugate-transpose operations are represented by superscripts $(\cdot)^{\rm T}$ and $(\cdot)^{\rm H}$, respectively; ${\rm tr}(\cdot)$ denotes the trace of a matrix; ${\mathbb C}^{a\times b}$ represents the set of complex $a\times b$ matrices; $|\cdot|$ and $\|\cdot\|$ denote the absolute value and the Euclidean norm, respectively; the Kronecker product is denoted by $\otimes$; ${\mathbb E}\{\cdot\}$ denotes statistical expectation; ${\rm diag}(a_1,a_2,\ldots,a_N)$ indicates a diagonal matrix with diagonal elements $\{a_i\}$. 

 

\section{System Model}\label{sec:2}

\begin{figure}[!t]
\centering
\includegraphics[width=0.5\textwidth]{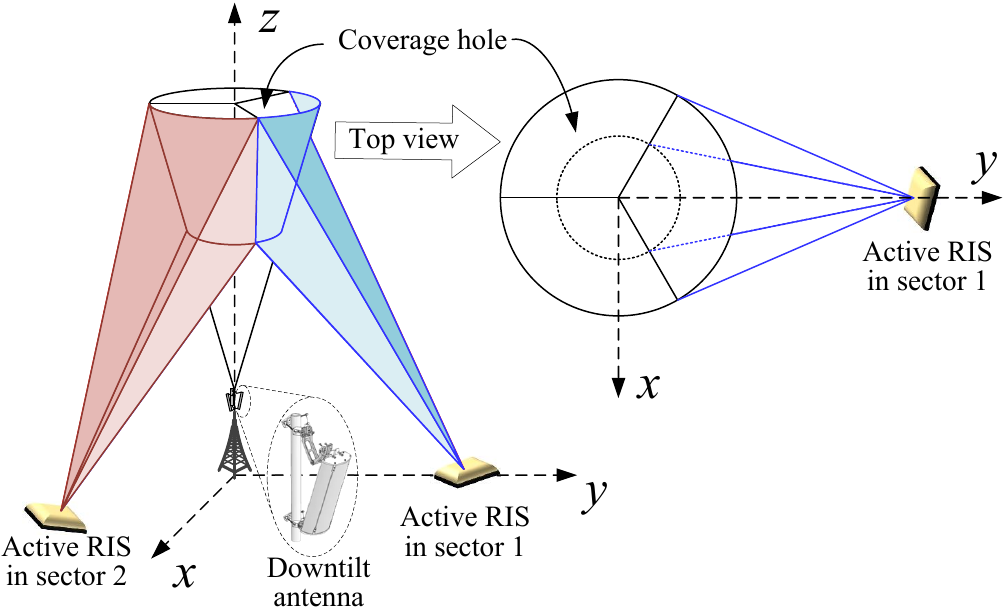}
\caption{Low-altitude coverage enhancement with multiple RISs.}
\label{Fig-system-model}
%\vspace{-0.5cm}
\end{figure}

We consider a cellular network as depicted in Fig. \ref{Fig-system-model}, where the BS is located at ${{\bf{p}}^{\rm{B}}} = {[ {0,0,{r_0}} ]^{\rm T}}$ in the Cartesian coordinate system and is equipped with a tri-sector antenna array. In each sector, the antenna array is a uniform planar array (UPA) with ${N^{\rm{B}}} = {N^{{\rm{B,v}}}} \times {N^{{\rm{B,h}}}}$ elements, covering a ${120^\circ}$ region in both vertical and horizontal directions. To provide superior wireless transmission services to terrestrial users, the BS antenna array is typically configured with a downtilt angle of ${\varpi ^{\rm{B}}}$.
As illustrated in Fig. \ref{Fig-system-model}, this downtilt configuration results in a vertical coverage gap directly above the BS, where the antenna gain is significantly reduced. This gap forms a three-dimensional frustum-shaped region\footnote{{UAVs locate in this region cannot establish a direct link with the BS, as they only ``see'' the rear face of the antenna array, where no electromagnetic radiation is emitted. This scenario is fundamentally different from those presented in \cite{R1-1} and \cite{R1-2}, where UAVs communicates with the sidelobe of the BS. Moreover, while \cite{R1-1} and \cite{R1-2} focus on interference suppression in passive RIS systems, we investigate active RIS scenarios and derive an upper bound on the noise power ratio between RISs and the BS. We will prove that, under practical parameter configurations, the RIS-introduced noise is negligible relative to that of the BS. This finding simplifies performance analyses and enables the extension of passive RIS frameworks to active RIS systems by ignore the RIS-induced noise.}}, referred to as the \emph{coverage hole} ${\mathcal V}$, which lies between the navigable altitudes $r_1$ and $r_2$. The coverage hole is approximately centered along the $z$-axis above the BS and characterized by a tilt angle of ${\varpi ^{\rm{B}}}+\pi/6$, resulting in a vertical extent of $R = {r_2} - {r_1}$. In the top view, this region appears as a circular blind spot with limited angular coverage across the three sectors. {To deliver enhanced uplink transmission for UAVs located in coverage holes without modifying existing wireless communication infrastructure, we mount active RISs on the rooftops of high-rise buildings with their apertures facing skyward. This setup constructs reflection links for each BS sector while eliminating adverse interference imposed by RIS sidelobes on ground users. The BS transmits limited control signaling to RISs over narrowband wireless control links to adjust the RIS reflection coefficients.} The RIS$_k$ ($k\in \{1,2,3\}$) in the $k$-th sector is located at ${\bf{p}}_k^{\rm{R}} = {[ {{x_k},{y_k},0} ]^{\rm T}}$ with an UPA of ${N^{\rm{R}}} = {N^{{\rm{R,v}}}} \times {N^{{\rm{R,h}}}}$ elements.

\subsection{Channel Model}

Considering the potential for a rich-scattering environment between the BS and RISs, the Rician model is adopted for the BS--RIS$_k$ channel ${{\bf{G}}_k} \in {{\mathbb C}^{{N^{\rm{B}}} \times {N^{\rm{R}}}}}$, which is expressed as ${{\bf{G}}_k} = {\bf{G}}_k^{\rm{L}} + {\bf{G}}_k^{\rm{N}}$,
where ${\bf{G}}_k^{\rm{L}}$ and ${\bf{G}}_k^{\rm{N}}$ are the LoS and non-LoS (NLoS) channel components, respectively. In the considered far-field, the LoS channel component is given by 
\begin{equation}\label{eq-G-kL}
	{\bf{G}}_k^{\rm{L}} = \sqrt {{\kappa }/({{\kappa  + 1}})} {\rho _k}{\bf{a}}_{k}^{\rm{BA}}{\bf{a}}_{k}^{{\rm{RD,H}}},
\end{equation}  
where $\kappa$ denotes the Rician factor and {the path loss $\rho_k=\frac{\lambda}{4\pi d_k}$ with $d_k$ being the distance between the BS and the RIS.} ${\bf a}_{k}^{\rm BA} = {\bf a}(\theta_{k}^{\rm BA},\phi_{k}^{\rm BA})$ is the array response vector (ARV) at the $k$-th sector of the BS, with $\theta_{k}^{\rm{BA}}$ and $\phi_{k}^{\rm{BA}}$ being the vertical and horizontal AoA of the LoS path, respectively. ${\bf{a}}_{k}^{{\rm{RD}}} = {\bf{a}}(\theta _{k}^{{\rm{RD}}},\phi _{k}^{{\rm{RD}}})$ is the ARV at RIS$_k$, with $\theta_{k}^{{\rm{RD}}}$ and $\phi_{k}^{{\rm{RD}}}$ being the vertical and horizontal AoD of the LoS path, respectively. {Considering the mobility patterns and trajectory variations of the UAV, we assume that AoDs of the LoS path vary across different channel coherence times, while remaining constant within a single coherence time.} For a UPA with adjacent element spacing of half a wavelength, the ARV is given by
\begin{equation}
	\begin{aligned}
		{\bf{a}}\left( {\theta ,\phi } \right) =& {\left[ {1,{e^{j\pi \cos\theta }}, \ldots ,{e^{j\left( {{N^{\rm{v}}} - 1} \right)\pi \cos\theta }}} \right]^{\rm{T}}} \\
		&\otimes {\left[ {1,{e^{j\pi\sin\theta\sin\phi }}, \ldots ,{e^{j\left( {{N^{\rm{h}}} - 1} \right)\pi\sin\theta\sin\phi  }}} \right]^{\rm{T}}},
	\end{aligned}
\end{equation}
where $\theta$ and $\phi$ indicate the angles of the signal propagation path, and $N^{\rm v}$ and $N^{\rm h}$ are the vertical and horizontal numbers of elements, respectively. A Rayleigh channel model incorporating the Rician factor and path loss is adopted for ${\bf G}_{k}^{\rm N}$, that is, $[{\bf G}_{k}^{\rm N}]_{m,n} \sim {\mathcal {CN}}(0,{\rho_k^2}/{(\kappa+1)})$.

For the UAV$_i$ located at ${{\bf{p}}_i} = {[ {{x_i},{y_i},{z_i}} ]^{\rm T}} \in {\mathcal V}$, the LoS link dominates the RIS$_k$--UAV$_i$ channel ${{\bf{g}}_{k,i}}\in {\mathbb C}^{N^{\rm R}\times 1}$, which can be modeled as
\begin{equation}\label{eq-gki}
{{\bf{g}}_{k,i}} = {\rho _{k,i}}{\bf{a}}_{k,i}^{{\rm{RA}}},
\end{equation}
where ${\rho _{k,i}}$ denotes the path loss and ${\bf{a}}_{k,i}^{{\rm{RA}}} = {\bf{a}}(\theta _{k,i}^{{\rm{RA}}},\phi _{k,i}^{{\rm{RA}}})$ is the ARV at RIS$_k$, with vertical and horizontal AoAs being $\theta _{k,i}^{{\rm{RA}}}$ and $\phi _{k,i}^{{\rm{RA}}}$, respectively. The cascaded channel from UAV$_i$ to the $k$-th antenna sector of the BS is then given by
\begin{equation}
	{{\bf{h}}_{k,i}} = {{\bf{G}}_k}{{\bf{\Gamma }}_k}{{\bf{g}}_{k,i}},
\end{equation}
where ${{\bm{\Gamma }}_k} = {\rm{diag}}( {{{\bm{\gamma }}_k}} ) \in {\mathbb C}^{N^{\rm R}\times N^{\rm R}}$ is the reflection matrix\footnote{{To facilitate theoretical analysis and {obtain} meaningful insights into the noise introduced by active RISs, we adopt a continuous phase shifter model in this work. Hardware non-idealities of RISs, including discrete phase shifts, mutual coupling \cite{PSM-MC}, angle-dependent phase shifts \cite{PSM-AD}, and amplitude-phase correlated phase shifts \cite{PSM-PD}, may lead to performance deviations in practical scenarios. However, these imperfections do not alter our fundamental conclusions, which are dominated by the macroscopic path loss characteristics among the BS, RIS and UAV. Furthermore, such issues can be alleviated via hardware optimization.}} of RIS$_k$. Here, ${{\bm{\gamma }}_k} = {\left[ {{\gamma _{k,1}}, \ldots ,{\gamma _{k,{N^{\rm{R}}}}}} \right]^{\rm{T}}}$ with ${{\gamma _{k,n}}}$ being the reflection coefficient of the $n$-th element in RIS$_k$. For active RISs, $1 \le | {{\gamma _{k,n}}} | \le {\gamma _{\max }}$, where ${\gamma _{\max }}$ is the maximum reflection amplitude of the RIS element. Note that the active RIS can be regarded as passive when ${\gamma _{\max }} = 1$.

\subsection{Signal Model}
As shown in Fig. \ref{Fig-system-model}, we assume that each BS sector can only receive signals reflected by the active RIS within that sector. Based on the considered channel model, the received signal at the $k$-th BS antenna sector can be expressed as \cite{ARIS-sig} 
\begin{equation}
	\begin{aligned}
		{{\bf{y}}_{k,i}} = {{\bf{G}}_k}{{\bf{\Gamma }}_k}\left( {{{\bf{g}}_{k,i}}x + {\bf{z}}_k^{\rm{R}}} \right) + {\bf{z}}_k^{\rm{B}} = {{\bf{h}}_{k,i}}x + {{\bf{G}}_k}{{\bf{\Gamma }}_k}{\bf{z}}_k^{\rm{R}} + {\bf{z}}_k^{\rm{B}},
	\end{aligned}
\end{equation}
where $x \sim {\mathcal {CN}}(0, {E_i})$ is the signal from UAV$_i$ with power $E_i$. ${\bf{z}}_k^{\rm{B}} \sim {\mathcal {CN}}(0, \sigma _{\rm{B}}^2{{\bf{I}}_{{N^{\rm{B}}} }})$ and ${\bf{z}}_k^{\rm{R}} \sim {\mathcal {CN}}(0, \sigma _{\rm{R}}^2{{\bf{I}}_{ {N^{\rm{R}}}}})$ are the noise terms at the $k$-th BS antenna sector and RIS$_k$, respectively. For typical passive RISs, $\sigma _{\rm{R}}^2 = 0$ is assumed. Denoting $E_{\rm R}$ as the maximum transmit power budget of the active RIS, the power constraint of RIS$_k$ is given by \cite{ARIS-power-1,ARIS-power-2} 
\begin{equation}\label{eq-RIS-power}
	{\mathbb E}\left\{{{{\left\| {{{\bf{\Gamma }}_k}\left( {{{\bf{g}}_{k,i}}x + {\bf{z}}_k^{\rm{R}}} \right)} \right\|}^2}}\right\} \le {E_{\rm{R}}}.
\end{equation}

Unlike passive RISs, active RISs not only amplify the reflected signal but also introduce additional noise from the amplification circuitry. The RIS-introduced noise ${{\bf{G}}_k}{{\bf{\Gamma }}_k}{\bf{z}}_k^{\rm{R}}$, combined with the BS-inherent noise ${\bf{z}}_k^{\rm{B}}$, collectively affects the received SNR. In the subsequent sections, we analyze the interaction between these noises. Building on this analysis, we explore the optimal reflection coefficient configuration for active RISs, aiming to maximize the received SNR and thereby enhance uplink transmission for UAVs located within the low-altitude coverage hole.

%%%%%%%%%%%%%%%%%%%%%%%%%%%%%%%%%%%%%%%%%%%%%%%%%%%%%%%%%%%%%%%%%%%%%%%%%%%%%%%%%%%%%%%%%%%%%%
\section{Analysis of Noise from RISs and BS}\label{sec:3}
In this section, we compare the noise power introduced by active RISs with the noise power at the BS to elucidate the impact of RIS-induced noise on the SNR. 
By collecting signals from three BS antenna sectors, we have
\begin{equation}
	{{\bf{y}}_i} = {\left[ {{\bf{y}}_{1,i}^{\rm{H}},{\bf{y}}_{2,i}^{\rm{H}},{\bf{y}}_{3,i}^{\rm{H}}} \right]^{\rm{H}}} = {{\bf{h}}_i}x + {{\bf{z}}^{\rm{R}}} + {{\bf{z}}^{\rm{B}}},
\end{equation}
where 
\begin{equation}
 {{\bf{h}}_i} = 
 \begin{bmatrix}
 {\bf{h}}_{1,i} \\
 {\bf{h}}_{2,i} \\
 {\bf{h}}_{3,i} 
 \end{bmatrix},~
{{\bf{z}}^{\rm{R}}} = 
\begin{bmatrix}
  {\bf{G}}_1 {\bf{\Gamma }}_1 {\bf{z}}_1^{{\rm{R}}} \\
  {\bf{G}}_2 {\bf{\Gamma }}_2 {\bf{z}}_2^{{\rm{R}}} \\
  {\bf{G}}_3 {\bf{\Gamma }}_3 {\bf{z}}_3^{{\rm{R}}}  
 \end{bmatrix},~
{{\bf{z}}^{\rm{B}}} = 
\begin{bmatrix}
 {\bf{z}}_1^{{\rm{B}}} \\
 {\bf{z}}_2^{{\rm{B}}} \\
 {\bf{z}}_3^{{\rm{B}}} 
 \end{bmatrix}.
\end{equation}
Here, ${\bf{h}}_i$ denotes the channel vector from UAV$_i$ to the BS.

In the context of low-altitude economies, UAVs need to transmit HD video data to the BS via high-capacity uplink. However, when UAVs are located in low-altitude coverage holes, the received SNR at the BS is significantly degraded. To enhance uplink transmission, we employ JS-MSC at the BS, leveraging diversity gain to improve the UAV's SNR. In JS-MSC, the BS combiner is designed as ${\bf{w}} = {{\bf{h}}_i}$ to jointly process signals received from different antenna sectors. In this case, the received signal can be expressed as 
\begin{equation}\label{eq-ri}
	r_i={{\bf{w}}^{\rm{H}}}{{\bf{y}}_i} ={{\bf{w}}^{\rm{H}}}{{\bf{h}}_i} x + {{\bf{w}}^{\rm{H}}}\left({{\bf{z}}^{\rm{R}}} + {{\bf{z}}^{\rm{B}}}\right).
\end{equation}
The average post-combining noise power at the BS is defined as ${\mathbb E}\{ | {{\bf{w}}^{\rm{H}}}{{\bf{z}}^{\rm{R}}} + {{\bf{w}}^{\rm{H}}}{{\bf{z}}^{\rm{B}}} |^2 \}$. Considering that $\{{\bf z}^{\rm R}_{k}\}_{k=1}^3$ in ${\bf z}^{\rm R}$ and $\{{\bf z}^{\rm B}_{k}\}_{k=1}^3$ in ${\bf z}^{\rm B}$ are independent Gaussian variables, we have 
\begin{equation}\label{eq-total-noise}
		{\mathbb E}\left\{ {{{\left| {{\bf{w}}^{\rm{H}}}\left({{\bf{z}}^{\rm{R}}} + {{\bf{z}}^{\rm{B}}}\right) \right|}^2}} \right\} = {\mathbb E}\left\{ {{{\left| { {{\bf{w}}^{\rm{H}}}{{\bf{z}}^{\rm{R}}}} \right|}^2}} \right\}+{\mathbb E}\left\{ {{{\left| {{{\bf{w}}^{\rm{H}}}{{\bf{z}}^{\rm{B}}}  } \right|}^2}} \right\}.
\end{equation}
In the following subsections, we derive closed-form expressions for ${\mathbb E}\{ | {{\bf{w}}^{\rm{H}}}{{\bf{z}}^{\rm{R}}} |^2 \}$ and ${\mathbb E}\{ | {{\bf{w}}^{\rm{H}}}{{\bf{z}}^{\rm{B}}} |^2 \}$ to obtain an upper bound on the noise power ratio between the BS and RISs.

\subsection{Closed-Form Expressions for Post-Combining Noise Power}
Since ${\bf{w}} = {{\bf{h}}_i}$ and ${\bf{z}}_k^{\rm{B}} \sim {\mathcal{CN}}(0, \sigma_{\rm{B}}^2{{\bf{I}}_{ N^{\rm{B}}}})$, the BS noise power can be expanded as ${\mathbb E}\{ | {\bf{w}}^{\rm{H}} {\bf{z}}^{\rm{B}} |^2 \} = \sigma_{\rm{B}}^2 {\mathbb E}\{ \| {\bf{h}}_i \|^2 \}$, and ${\mathbb E}\{ \| {\bf{h}}_i \|^2 \}$ has a closed-form expression given in \emph{Lemma} \ref{Lemma-1}. 

\begin{lemma}\label{Lemma-1}
	In the considered multi-RIS-assisted low-altitude communication with a tri-sector BS, the average power gain of the channel can be expressed as 
	\begin{equation}
		{\mathbb E}\left\{ {{{\left\| {{{\bf{h}}_i}} \right\|}^2}} \right\} = {{{N^{\rm{B}}}}}/{{(\kappa  + 1)}}\sum\nolimits_{k = 1}^3 {\rho _{k,i}^2\rho _k^2F_{0,k,i}} ,
	\end{equation}
where $F_{0,k,i}={\kappa {{| {{\bm{\gamma }}_k^{\rm{H}}{\bf{a}}_{k,i}^{\rm{R}}} |}^2} + {{\| {{{\bm{\gamma }}_k}} \|}^2}}$ and ${\bf{a}}_{k,i}^{\rm{R}} = ({\bf{a}}_{k,i}^{{\rm{RA}}})^* \odot {\bf{a}}_{k}^{{\rm{RD}}}$.
\end{lemma}
\begin{IEEEproof}
	Please refer to Appendix \ref{APP-L1}.
\end{IEEEproof}

By expanding ${\bf{h}}_i$ and ${\bf{z}}^{\rm{R}}$, the RIS-introduced noise power ${\mathbb E}\{ | {\bf{w}}^{\rm{H}} {\bf{z}}^{\rm{R}} |^2 \}$ can be expressed as 
\begin{equation}\label{eq-EwHz}
	\begin{aligned}
		&{\mathbb E}\left\{ {{{\left| {{{\bf{w}}^{\rm{H}}}{{\bf{z}}^{\rm{R}}}} \right|}^2}} \right\} \\
		&= \sum\limits_{k = 1}^3 {\sum\limits_{k' = 1}^3 {\mathbb E}{\left\{ {{\bf{g}}_{k,i}^{\rm{H}}{\bf{\Gamma }}_k^{\rm{H}}{\bf{G}}_k^{\rm{H}}{{\bf{G}}_k}{{\bf{\Gamma }}_k}{\bf{z}}_k^{\rm{R}}{\bf{z}}_{k'}^{{\rm{R,H}}}{\bf{\Gamma }}_{k'}^{\rm{H}}{\bf{G}}_{k'}^{\rm{H}}{{\bf{G}}_{k'}}{{\bf{\Gamma }}_{k'}}{{\bf{g}}_{k',i}}} \right\}} } .
	\end{aligned}
\end{equation}
Since ${\bf{z}}_k^{\rm{R}}$ and ${\bf{z}}_{k'}^{\rm{R}}$ are independent Gaussian vectors for $k \ne k'$, we have ${\mathbb E}\{ {\bf{z}}_k^{\rm{R}} {\bf{z}}_k^{\rm{R,H}} \} = \sigma_{\rm{R}}^2 {\bf{I}}_{N^{\rm{R}}}$ and ${\mathbb E}\{ {\bf{z}}_k^{\rm{R}} {\bf{z}}_{k'}^{\rm{R,H}} \} = {\bf{0}}$ for $k \ne k'$. Therefore, \eqref{eq-EwHz} simplifies to
	\begin{align}
		&{\mathbb E}\left\{ {{{\left| {{{\bf{w}}^{\rm{H}}}{{\bf{z}}^{\rm{R}}}} \right|}^2}} \right\} \notag \\
		 &= \sigma _{\rm{R}}^2\sum\nolimits_{k = 1}^3 {\mathbb E}{\left\{ {{\bf{g}}_{k,i}^{\rm{H}}{\bf{\Gamma }}_k^{\rm{H}}{\bf{G}}_k^{\rm{H}}{{\bf{G}}_k}{{\bf{\Gamma }}_k}{\bf{\Gamma }}_k^{\rm{H}}{\bf{G}}_k^{\rm{H}}{{\bf{G}}_k}{{\bf{\Gamma }}_k}{{\bf{g}}_{k,i}}} \right\}} \notag \\
		& = \sigma _{\rm{R}}^2\sum\nolimits_{k = 1}^3 {{\bf{g}}_{k,i}^{\rm{H}}{\bf{\Gamma }}_k^{\rm{H}}{\mathbb E}\left\{ {{\bf{G}}_k^{\rm{H}}{{\bf{G}}_k}{{\bf{\Gamma }}_k}{\bf{\Gamma }}_k^{\rm{H}}{\bf{G}}_k^{\rm{H}}{{\bf{G}}_k}} \right\}} {{\bf{\Gamma }}_k}{{\bf{g}}_{k,i}}, \label{eq-EwHzR}
	\end{align}
where the second equality holds because both ${\bf{g}}_{k,i}$ and ${\bf{\Gamma}}_k$ are long-term parameters.
A closed-form expression for ${\mathbb E}\left\{ {\bf{G}}_k^{\rm{H}} {\bf{G}}_k {\bf{\Gamma}}_k {\bf{\Gamma}}_k^{\rm{H}} {\bf{G}}_k^{\rm{H}} {\bf{G}}_k \right\}$ is provided in Lemma \ref{Lemma-2}.

\begin{lemma}\label{Lemma-2}
	In the considered scenario where the BS--RIS$_k$ channel ${\bf G}_k$ is composed of a statistical LoS link and a Rayleigh NLoS component, the fourth-order moment of ${\bf G}_k$ incorporating ${\bm \Gamma_k}$ can be given by
	\begin{equation}\label{eq-L2}
	\begin{aligned}
		&{\mathbb E}\left\{ {{\bf{G}}_k^{\rm{H}}{{\bf{G}}_k}{{\bf{\Gamma }}_k}{\bf{\Gamma }}_k^{\rm{H}}{\bf{G}}_k^{\rm{H}}{{\bf{G}}_k}} \right\} ={\bf{G}}_k^{{\rm{L}},{\rm{H}}}{\bf{G}}_k^{\rm{L}}{\left| {{{\bf{\Gamma }}_k}} \right|^2}{\bf{G}}_k^{{\rm{L}},{\rm{H}}}{\bf{G}}_k^{\rm{L}} \\
		&+ \frac{{\rho _k^2}}{{\kappa  + 1}}\left( {2{N^{\rm{B}}}{\mathop{\rm Re}\nolimits} {\left( {{\bf{G}}_k^{{\rm{L}},{\rm{H}}}{\bf{G}}_k^{\rm{L}}{{\left| {{{\bf{\Gamma }}_k}} \right|}^2}} \right)} + {{\left\| {{{\bm{\gamma }}_k}} \right\|}^2}{\bf{G}}_k^{{\rm{L}},{\rm{H}}}{\bf{G}}_k^{\rm{L}}} \right)\\
		& + \frac{{\rho _k^2}}{{\kappa  + 1}}{\left\| {{\bf{G}}_k^{\rm{L}}{{\bf{\Gamma }}_k}} \right\|^2}{{\bf{I}}_{{N^{\rm{R}}}}} + \frac{{{N^{\rm{B}}}\rho _k^4}}{{{{(\kappa  + 1)}^2}}}\left( {{N^{\rm{B}}}|{{\bf{\Gamma }}_k}{|^2} + {{\left\| {{{\bm{\gamma }}_k}} \right\|}^2}{{\bf{I}}_{{N^{\rm{R}}}}}} \right).
		 	\end{aligned}
	\end{equation}
\end{lemma}
\begin{IEEEproof}
	Please refer to Appendix \ref{APP-L2}.
\end{IEEEproof}

By leveraging Lemma \ref{Lemma-2}, a closed-form expression for the RIS-introduced noise power is presented in Theorem \ref{Theo-1}.
 
\begin{theorem}\label{Theo-1}
	In the considered multi-RIS-assisted low-altitude communication with a tri-sector BS, when JS-MSC is employed at the BS, the average noise power introduced by active RISs can be expressed as 
	\begin{equation}\label{eq-T1}
		{\mathbb E}\left\{ {{{\left| {{{\bf{w}}^{\rm{H}}}{{\bf{z}}^{\rm{R}}}} \right|}^2}} \right\} =\frac{{\sigma _{\rm{R}}^2{N^{\rm{B}}}}}{{{{\left(\kappa  + 1\right)}^2}}}\sum\limits_{k = 1}^3 {\rho _k^4\rho _{k,i}^2\left( {{F_{1,k,i}} + {F_{2,k,i}} + {F_{3,k}}} \right)} ,
	\end{equation}
where 
\begin{subequations} 
	\begin{align}
		{F_{1,k,i}} &= \kappa \left( {{N^{\rm{B}}}{\kappa} + 1 } \right){\left| {{\bm{\gamma }}_k^{\rm{H}}{\bf{a}}_{k,i}^{\rm{R}}} \right|^2}{\left\| {{{\bm{\gamma }}_k}} \right\|^2},\\
		{F_{2,k,i}} &= 2{N^{\rm{B}}}\kappa {\mathop{\rm Re}\nolimits} {\left( {{\bm{\gamma }}_k^{\rm{H}}{\bf{a}}_{k,i}^{\rm{R}}{\bf{a}}_{k,i}^{{\rm{R}},{\rm{H}}}\left( {{\bm{\gamma }}_k^{\rm{*}} \odot {{\bm{\gamma }}_k} \odot {{\bm{\gamma }}_k}} \right)} \right)},\\
		{F_{3,k}} &= \left( {\kappa  + 1} \right){\left\| {{{\bm{\gamma }}_k}} \right\|^4} + {N^{\rm{B}}}{\rm{tr}}\left( {|{{\bf{\Gamma }}_k}{|^4}} \right).
	\end{align} 
\end{subequations}
\end{theorem}
\begin{IEEEproof}
    By substituting \eqref{eq-G-kL}, \eqref{eq-gki}, and \eqref{eq-L2} into \eqref{eq-EwHzR} and performing algebraic simplifications, \eqref{eq-T1} is obtained. Detailed steps are omitted due to space limitations.
\end{IEEEproof}

{From a large-scale path loss perspective, the key insight gleaned from Lemma \ref{Lemma-1} and Theorem \ref{Theo-1} collectively is that RIS-induced noise includes additional path loss relative to BS-inherent noise. This insight implies the potential negligibility of RIS-induced noise for SNR analysis. To elucidate the practical implications of reflection-dependent active RIS noise, we further derive an upper bound on the ratio of RIS-induced noise to BS-inherent noise.}

\subsection{Upper Bound for Noise Power Ratio}

Based on Lemma \ref{Lemma-1} and Theorem \ref{Theo-1}, the total noise power in \eqref{eq-total-noise} can be expressed as 
\begin{equation}
		\begin{aligned}
			&{\mathbb E}\left\{ {{{\left| {{\bf{w}}^{\rm{H}}}\left({{\bf{z}}^{\rm{R}}} + {{\bf{z}}^{\rm{B}}}\right) \right|}^2}} \right\}=\frac{{{N^{\rm{B}}}{\sigma ^2}}}{{\kappa  + 1}}\times\\
			&\sum\limits_{k = 1}^3 {\rho _{k,i}^2\rho _k^2\left( {\underbrace {{F_{0,k,i}}}_{{\rm{BS - inherent}}} + \underbrace {\frac{{\rho _k^2\left( {{F_{1,k,i}} + {F_{2,k,i}} + {F_{3,k}}} \right)}}{{\kappa  + 1}}}_{{\rm{RIS - introduced}}}} \right)}  ,
		\end{aligned}
\end{equation}
where $\sigma_{\rm{B}}^2 = \sigma_{\rm{R}}^2 = \sigma^2$ is assumed for simplicity. To compare the BS-inherent and RIS-introduced noise, we define the noise power ratio as 
\begin{equation}
	{\beta _{k,i}} = \frac{{\rho _k^2\left( {{F_{1,k,i}} + {F_{2,k,i}} + {F_{3,k}}} \right)}}{{\left( {\kappa  + 1} \right){F_{0,k,i}}}}.
\end{equation}

In a given system, this noise power ratio depends on the specific reflection amplitude and phase of the RIS. According to the constraint on reflection amplitude $1 \le | \gamma_{k,n} | \le \gamma_{\max}$, we have the following inequalities:   
\begin{subequations}
 \begin{align} 
	 {N^{\rm{R}}} \le {\left\| {{{\bm{\gamma }}_k}} \right\|^2} &\le {N^{\rm{R}}}\gamma _{\max }^2, \label{eq-gamma1_2}\\
	\left|{\bf{a}}_{k,i}^{{\rm{R}},{\rm{H}}}\left( {{\bm{\gamma }}_k^{\rm{*}} \odot {{\bm{\gamma }}_k} \odot {{\bm{\gamma }}_k}} \right)\right| &\le {N^{\rm{R}}}\gamma _{\max }^3, \label{eq-gamma3}\\
	{\rm{tr}}\left( {|{{\bf{\Gamma }}_k}{|^4}} \right) &\le {N^{\rm{R}}}\gamma _{\max }^4. \label{eq-gamma4} 
    \end{align}  
\end{subequations} 
By applying \eqref{eq-gamma1_2}–\eqref{eq-gamma4} to $F_{0,k,i}$, $F_{1,k,i}$, $F_{2,k,i}$, and $F_{3,k}$, we obtain the following bounds: 
\begin{subequations} 
 \begin{align} 
			{F_{0,k,i}} &\ge F_{0,k,i}^{{\rm{Low}}} \buildrel \Delta \over = \kappa {\left| {{\bm{\gamma }}_k^{\rm{H}}{\bf{a}}_{k,i}^{\rm{R}}} \right|^2} + {N^{\rm{R}}},\\
			{F_{1,k,i}}& \le F_{1,k,i}^{{\rm{Upp}}} \buildrel \Delta \over = \left( {{N^{\rm{B}}}{\kappa ^2} + \kappa } \right){\left| {{\bm{\gamma }}_k^{\rm{H}}{\bf{a}}_{k,i}^{\rm{R}}} \right|^2}{N^{\rm{R}}}\gamma _{\max }^2,\\
			{F_{2,k,i}}& \le F_{2,k,i}^{{\rm{Upp}}} \buildrel \Delta \over = 2{N^{\rm{B}}}\kappa {N^{\rm{R}}}\gamma _{\max }^3\left| {{\bm{\gamma }}_k^{\rm{H}}{\bf{a}}_{k,i}^{\rm{R}}} \right|,\\
			{F_{3,k}} &\le F_{3,k}^{{\rm{Upp}}} \buildrel \Delta \over = \left( {\kappa  + 1} \right){N^{{\rm{R,2}}}}\gamma _{\max }^4 + {N^{\rm{B}}}{N^{\rm{R}}}\gamma _{\max }^4.
    \end{align}  
\end{subequations}
Based on these bounds, the noise power ratio satisfies
\begin{equation}
	{\beta _{k,i}} \le \frac{{\rho _k^2\left( {{F_{1,k,i}^{{\rm{Upp}}}} + {F_{2,k,i}^{{\rm{Upp}}}} + {F_{3,k}^{{\rm{Upp}}}}} \right)}}{{\left( {\kappa  + 1} \right) F_{0,k,i}^{{\rm{Low}}} }}=\frac{{{N^{\rm{R}}}\rho _k^2\gamma _{\max }^2}}{{\left( {\kappa  + 1} \right)}}f\left( x \right),
\end{equation}
where $x = | {{\bm{\gamma }}_k^{\rm{H}}{\bf{a}}_{k,i}^{\rm{R}}} | \in [ {0,{N^{\rm{R}}}{\gamma _{\max }}} ]$ and 
\begin{equation}\label{eq-fx-def}
	\begin{aligned}
		f\left( x \right) =& \frac{{\left( {{N^{\rm{B}}}{\kappa ^2} + \kappa } \right){x^2} + 2{N^{\rm{B}}}\kappa {\gamma _{\max }}x}}{{\kappa {x^2} + {N^{\rm{R}}}}} \\
		&+ \frac{{\kappa {N^{\rm{R}}} + {N^{\rm{R}}} + {N^{\rm{B}}}}}{{\kappa {x^2} + {N^{\rm{R}}}}}\gamma _{\max }^2.
	\end{aligned}
\end{equation}
Based on the proof in Appendix \ref{App-fx}, the maximum of $f(x)$ is given by
\begin{equation}\label{eq-fxopt}
	f\left( {{x_{{\rm{opt}}}}} \right) = {a_1} + \frac{{a_2^2\sqrt {{\kappa ^2}a_3^2 + \kappa a_2^2{N^{\rm{R}}}} }}{{{{\left( {\sqrt {{\kappa ^2}a_3^2 + \kappa a_2^2{N^{\rm{R}}}}  - \kappa {a_3}} \right)}^2} + {N^{\rm{R}}}\kappa a_2^2}},
\end{equation}
where $a_1 = N^{\rm{B}} \kappa + 1$, $a_2 = 2 N^{\rm{B}} \kappa \gamma_{\max}$, and $a_3 = (N^{\rm{B}} + N^{\rm{R}} + \kappa N^{\rm{R}}) \gamma_{\max}^2 - a_1 N^{\rm{R}}$. 

Thus, an upper bound of $\beta_{k,i}$ under arbitrary RIS configuration is 
\begin{equation}\label{eq-beta1}
	{\beta _{k,i}^{{\rm Upp},1}}=f\left( x_{\rm opt} \right){{{N^{\rm{R}}}\rho _k^2\gamma _{\max }^2}}/{{\left( {\kappa  + 1} \right)}}.
\end{equation}
{Note that $\beta_{k,i}^{\rm{Upp},1}$ is derived under the reflection amplitude constraint $1 \le | \gamma_{k,n} | \le \gamma_{\max}$. Hence, when $\gamma_{\max} \to 1$, the bound is tight. When $\gamma_{\max}$ increases, the bound becomes looser. Since $\beta_{k,i}^{\rm{Upp},1}$ is independent of the specific RIS reflection configuration, it can serve as a metric to evaluate the negligibility of the RIS-induced noise. When $\beta_{k,i}^{\rm{Upp},1}$ is sufficiently small, RIS-induced noise is far smaller than the BS-inherent noise regardless of the RIS configuration, and thus can be reasonably neglected.} However, $\beta_{k,i}^{\rm{Upp},1}$ is still complex and may not provide intuitive insights. To resolve this, a more tractable upper bound is presented in Theorem \ref{Theo-betaU2}, with a trade-off in looseness.

\begin{theorem}\label{Theo-betaU2} 
    For a tri-sector BS employing JS-MSC, the ratio between the RIS-introduced and BS-inherent noise power in the link cascaded by the active RIS$_k$ is upper bounded by $\beta_{k,i}^{\rm{Upp},2}$ as given in \eqref{eq-beta2}. 

	\begin{figure*}
		\begin{equation}\label{eq-beta2}
			\beta _{k,i}^{{\rm{Upp}},2}=\left\{ \begin{aligned}
				&\frac{\rho _k^2\gamma _{\max }^2}{\kappa+1}\left( {2{N^{\rm{B}}}\sqrt {\kappa {N^{\rm{R}}}} {\gamma _{\max }} + 2{N^{\rm{R}}}\left( {{N^{\rm{B}}}\kappa  + 1} \right)} \right), & &\gamma _{\max }^2 \le \frac{{\left( {{N^{\rm{B}}}\kappa  + 1} \right){N^{\rm{R}}}}}{{{N^{\rm{B}}} + {N^{\rm{R}}} + \kappa {N^{\rm{R}}}}},\\
				&\frac{\rho _k^2\gamma _{\max }^3}{\kappa+1}\left( {2{N^{\rm{B}}}\sqrt {\kappa {N^{\rm{R}}}}  + {\gamma _{\max }}\left( {{N^{\rm{R}}}\kappa +{N^{\rm{R}}} + {N^{\rm{B}}}} \right)} \right),  & &\gamma _{\max }^2 > \frac{{\left( {{N^{\rm{B}}}\kappa  + 1} \right){N^{\rm{R}}}}}{{{N^{\rm{B}}} + {N^{\rm{R}}} + \kappa {N^{\rm{R}}}}}.
			\end{aligned} \right.
		\end{equation}
		%\vspace{-0.5cm}
	\hrulefill
	\end{figure*}
\end{theorem}
\begin{IEEEproof}
	Please refer to Appendix \ref{APP-T2}.
\end{IEEEproof}

{Theorem \ref{Theo-betaU2} offers useful insight into the relationship between RIS-introduced and BS-inherent noise, which can be applied to simplify RIS optimization.} In a typical 5G system operating at 3.55 GHz ($N^{\rm{B}} = 12 \times 8$, $N^{\rm{R}} = 10 \times 10$, $\gamma_{\max} = 15$, $\kappa = 1$, and a BS-RIS distance of 100 meters), we have $\beta_{k,i}^{\rm{Upp},2} = 0.0233$, {indicating that the intractable RIS-introduced noise can be ignored.} Since $\beta_{k,i}^{\rm{Upp},2}$ is linearly proportional to $N^{\rm{B}}$ and $N^{\rm{R}}$, smaller antenna arrays lead to even smaller ratios. As RISs are deployed further from the BS, path loss increases, which reduces $\beta_{k,i}^{\rm{Upp},2}$. {Therefore, from the perspective of whether active RIS noise can be neglected, active RISs should not be deployed in close proximity to the receiver, which reverses the deployment strategy adopted for noise-free passive RISs.}

{Importantly, at the RIS side, $\beta_{k,i}^{\rm{Upp},2}$ is more sensitive to the reflection amplitude $\gamma_{\max}$ than to the RIS size $N^{\rm R}$.} When $\gamma_{\max}$ is small, the ratio is proportional to $\gamma_{\max}^3$. {Notably, when $\gamma_{\max} \to 1$ , $\beta_{k,i}^{\rm{Upp},2}$ drops below $10^{-4}$.} In contrast, when $\gamma_{\max}$ is large, for example $\gamma_{\max} = 15$ as in the example system, and satisfies $\gamma_{\max}^2 > \frac{ (N^{\rm{B}} \kappa + 1) N^{\rm{R}} }{ N^{\rm{B}} + N^{\rm{R}} + \kappa N^{\rm{R}} }$, then the ratio becomes proportional to $\gamma_{\max}^4$. Adjusting parameters $N^{\rm R}$, $N^{\rm B}$, $\rho_k$, and $\gamma_{\max}$ to reduce $\beta_{k,i}^{\rm{Upp},2}$ also results in a decrease in the SNR because of the reduced received signal power, as will be detailed in the end of Sec. \ref{sec:4}. When these parameters are chosen such that $\beta_{k,i}^{\rm{Upp},2}$ is sufficient small, the noise introduced by RISs becomes negligible compared to the inherent noise at the BS. In this case, the amplification capability of active RIS always provide positive gains.

 {Based on aforementioned insights, in practical systems where the loose noise upper bound $\beta _{k,i}^{{\rm{Upp}},2}$ is smaller than an acceptable noise ratio $\tau$, the RIS-introduced noise can be safely neglected: 
\begin{equation}
	{\mathbb E}\left\{ {{{\left| {{\bf{w}}^{\rm{H}}}\left({{\bf{z}}^{\rm{R}}} + {{\bf{z}}^{\rm{B}}}\right) \right|}^2}} \right\} \approx{\mathbb E}\left\{ {{{\left| {{{\bf{w}}^{\rm{H}}}{{\bf{z}}^{\rm{B}}}  } \right|}^2}} \right\}.
\end{equation}
This approximation simplifies performance analyses and transmission design by enabling the  extension of passive RIS frameworks to active RIS systems without considering the RIS-induced noise. By setting $\beta_{k,i}^{\rm{Upp},2}=\tau$, we can establish the relationships among multiple parameters, including $\gamma_{\max}$, $N^{\rm{R}}$, $N^{\rm{B}}$, and $d_k$ (or $\rho_k$). For RIS deployment in existing cellular infrastructure, this relationship can guide system designers to determine the maximum feasible amplification gain and array scale, as well as the minimum distance between the RIS and receiver, beyond which the noise simplification no longer holds.}
 
\section{Reflection Design for Uplink Enhancement}\label{sec:4}

To enhance uplink transmission, we derive an approximate lower bound of the average received SNR in this section. {The RIS reflection design for maximizing this approximate lower bound is decoupled into two sub-problems: reflection phase design and reflection amplitude design. Optimal closed-form solutions are subsequently developed to solve these two sub-problems in sequence. Furthermore, the performance of the active RIS is evaluated by analyzing the tradeoff between the array size and reflection amplitude.}
 
Given the received signal in \eqref{eq-ri}, the average received SNR is defined as 
\begin{equation}\label{eq-E-SNR}
	{\eta_i^{\rm ave}}={\mathbb E}\left\{\frac{{{E_i}{{\left\| {{{\bf{h}}_i}} \right\|}^4}}}{{{{\left| {{{\bf{w}}^{\rm{H}}}\left( {{{\bf{z}}^{\rm{R}}} + {{\bf{z}}^{\rm{B}}}} \right)} \right|}^2}}}\right\}.
\end{equation}
Considering that ${\eta_i^{\rm LB}}={E_i}{\mathbb E}\{{{{{| {{\bf{w}}^{\rm{H}}}{{\bf{h}}_i}|}^2}}}\}/{\mathbb E}\{{{{{| {{{\bf{w}}^{\rm{H}}}( {{{\bf{z}}^{\rm{R}}} + {{\bf{z}}^{\rm{B}}}} )} |}^2}}}\}$ is a lower bound of ${\eta_i^{\rm ave}}$ \cite{LB-SNR} and ${\mathbb E}\{ {{{| {{\bf{w}}^{\rm{H}}}\left({{\bf{z}}^{\rm{R}}} + {{\bf{z}}^{\rm{B}}}\right) |}^2}} \} \approx{\mathbb E}\{ {{{| {{{\bf{w}}^{\rm{H}}}{{\bf{z}}^{\rm{B}}}  } |}^2}} \}$, the approximate lower bound of \eqref{eq-E-SNR} can be written as\footnote{We consider typical setups where the analytical noise power ratio upper bound $	{\beta _{k,i}^{{\rm Upp},1}}$ is sufficiently small such that the RIS-introduced noise can be neglected.}
\begin{equation}
	{\eta _i} = {{{E_i}{\mathbb E}\left\{ {{{\left\| {{{\bf{h}}_i}} \right\|}^4}} \right\}}}/{{\mathbb E}\left\{ {{{\left| {{\bf{w}}^{\rm{H}}} {{\bf{z}}^{\rm{B}}} \right|}^2}} \right\}}.
\end{equation}

The RIS reflection phase and amplitude in ${\bf h}_i$ and ${\bf w}$ jointly determines ${\bf \eta}_i$. To derive a closed-form expression of $\eta_i$, we present the following lemma regarding the average received signal power. 
\begin{lemma}\label{L-Rec}
    In the context of multi-RIS-assisted low-altitude communications with a tri-sector BS, when JS-MSC is employed, the average received signal power is given by 
	\begin{equation}
			\begin{aligned}
				&{\mathbb E}\left\{ {{{\left\| {{{\bf{h}}_i}} \right\|}^4}} \right\}\\
				=&\frac{{{N^{\rm{B}}}}}{{{{\left( {\kappa  + 1} \right)}^2}}}\sum\nolimits_{k = 1}^3 {\rho _{k,i}^4\rho _k^4\left( {\left( {{N^{\rm{B}}} + 1} \right)F_{0,k,i}^2 - {\kappa^2}{{\left| {{\bm{\gamma }}_k^{\rm{H}}{\bf{a}}_{k,i}^{\rm{R}}} \right|}^4}} \right)} \\
				&+ \frac{{{N^{{\rm{B,2}}}}}}{{{{\left( {\kappa  + 1} \right)}^2}}}\sum\nolimits_{k = 1}^3 {\rho _{k,i}^2\rho _k^2{F_{0,k,i}}\sum\limits_{m \ne k}^3 {\rho _{m,i}^2\rho _m^2{F_{0,m,i}}} } .
			\end{aligned}
	\end{equation}
\end{lemma}
\begin{IEEEproof}
	Please refer to Appendix \ref{APP-L-Rec}.
\end{IEEEproof}

Based on Lemmas \ref{Lemma-1} and \ref{L-Rec}, which provide closed-form expressions for ${\mathbb E}\left\{ \left\| {\bf{h}}_i \right\|^4 \right\}$ and ${\mathbb E}\left\{ \left\| {\bf{h}}_i \right\|^2 \right\}$, the approximate lower bound of the received SNR is given by
\begin{equation}
	{\eta _i} = \frac{{{E_i}}}{\sigma_{\rm{B}}^2} {{{\mathbb E}\left\{ {{{\left\| {{{\bf{h}}_i}} \right\|}^4}} \right\}}}/{{{\mathbb E}\left\{ {{{\left\| {{{\bf{h}}_i}} \right\|}^2}} \right\}}}.
\end{equation}

{Note that $\eta_i$ is a complex function of $\{ {\bm \gamma_k} \}_{k=1}^{3}$, which renders the direct maximization of $\eta_i$ intractable. To address this challenge, we decouple the reflection vector as ${\bm{\gamma}}_k =   {\bm{\omega}}_k \odot {\bm{\alpha}}_k$, and sequentially derive closed-form solutions for ${\bm{\omega}}_k$ and ${\bm{\alpha}}_k$, respectively.}  Here,
\begin{subequations} 
 \begin{align} 
  {\bm{\omega}}_k &= \left[ e^{j \angle \gamma_{k,1}}, \ldots, e^{j \angle \gamma_{k,N^{\rm{R}}}} \right]^{\rm{T}}, \\
  {\bm{\alpha}}_k &= \left[ | \gamma_{k,1} |, \ldots, | \gamma_{k,N^{\rm{R}}} | \right]^{\rm{T}}, 
 \end{align} 
\end{subequations} 
denote the reflection phase and amplitude vectors, respectively.
 
\subsection{Reflection Phase Design}

{When $\{{\bm \alpha_k}\}_{k=1}^{3}$ are fixed, ${{\| {{{\bm{\gamma }}_k}} \|}^2}={{\| {{{\bm{\alpha }}_k}} \|}^2}$ in $F_{0,k,i}$ is a constant. Thus, we can simplify $\eta_i$ as a function} of $x_k = | {\bm{\gamma}}_k^{\rm{H}} {\bf{a}}_{k,i}^{\rm{R}} |^2 \in [0, \left\| {\bm{\alpha}}_k \right\|_1^2]$, as follows: 
\begin{multline} 
	{\eta _i} = \frac{{{N^{\rm{B}}}{E_i}}}{{\left( {\kappa  + 1} \right)\sigma _{\rm{B}}^2}}\sum\nolimits_{k = 1}^3 {\rho _{k,i}^2\rho _k^2\left( {\kappa x_k + {{\left\| {{{\bm{\alpha }}_k}} \right\|}^2}} \right)} \\
	+ \frac{{2\kappa {E_i}\sum\nolimits_{k = 1}^3 {\rho _{k,i}^4\rho _k^4{{\left\| {{{\bm{\alpha }}_k}} \right\|}^2}x_k}  + \sum\nolimits_{k = 1}^3 {\rho _{k,i}^4\rho _k^4{{\left\| {{{\bm{\alpha }}_k}} \right\|}^4}} }}{{\left( {\kappa  + 1} \right)\sigma _{\rm{B}}^2\sum\nolimits_{k = 1}^3 {\rho _{k,i}^2\rho _k^2\left( {\kappa x_k + {{\left\| {{{\bm{\alpha }}_k}} \right\|}^2}} \right)} }}. 
\end{multline}
The derivative of $\eta_i$ with respect to $x_m$ is given by 
\begin{equation}
	\frac{{\partial {\eta _i}}}{{\partial {x_m}}}=\frac{{\kappa \rho _{m,i}^2\rho _m^2{E_i}}}{{\left( {\kappa  + 1} \right)\sigma _{\rm{B}}^2}}\frac{{\sum\nolimits_{k = 1}^3 {\rho _{k,i}^2\rho _k^2{X_{k,m,i}}} }}{{{{\left( {\sum\nolimits_{k = 1}^3 {\rho _{k,i}^2\rho _k^2\left( {\kappa x_k + {{\left\| {{{\bm{\alpha }}_k}} \right\|}^2}} \right)} } \right)}^2}}},
\end{equation}
where 
\begin{multline} 
		{X_{k,m,i}} = \left( {\kappa x_k + {{\left\| {{{\bm{\alpha }}_k}} \right\|}^2}} \right)\sum\limits_{n = 1}^3 {\rho _{n,i}^2\rho _n^2\left( {{N^{\rm{B}}}\kappa x_k + {N^{\rm{B}}}{{\left\| {{{\bm{\alpha }}_n}} \right\|}^2}} \right)}  \\
		- {\left\| {{{\bm{\alpha }}_k}} \right\|^2}\rho _{k,i}^2\rho _k^2\left( {2\kappa {{\left| {{\bm{\gamma }}_k^{\rm{H}}{\bf{a}}_{k,i}^{\rm{R}}} \right|}^2} + {{\left\| {{{\bm{\alpha }}_k}} \right\|}^2}} \right)\\
		+ 2\rho _{m,i}^2\rho _m^2{\left\| {{{\bm{\alpha }}_m}} \right\|^2}\left({\left\| {{{\bm{\alpha }}_k}} \right\|^2} + \kappa {\left| {{\bm{\gamma }}_k^{\rm{H}}{\bf{a}}_{k,i}^{\rm{R}}} \right|^2} \right).
\end{multline}
Since $\kappa {| {{\bm{\gamma }}_k^{\rm{H}}{\bf{a}}_{k,i}^{\rm{R}}} |^2} + {\| {{{\bm{\alpha }}_k}} \|^2} \ge {\| {{{\bm{\alpha }}_k}} \|^2}$ and ${N^{\rm{B}}}\kappa {| {{\bm{\gamma }}_k^{\rm{H}}{\bf{a}}_{k,i}^{\rm{R}}} |^2} + {N^{\rm{B}}}{\| {{{\bm{\alpha }}_k}} \|^2} \ge 2\kappa {| {{\bm{\gamma }}_k^{\rm{H}}{\bf{a}}_{k,i}^{\rm{R}}} |^2} + {\| {{{\bm{\alpha }}_k}} \|^2}$, it follows that $X_{k,m,i} > 0$. In other words, ${\partial \eta_i}/{\partial x_m} > 0$, and thus $\eta_i$ increases with $x_m$. Therefore, $\eta_i$ is maximized when $x_m = \left\| {\bm{\alpha}}_m \right\|_1^2$, indicating that the optimal phase of RIS$_m$ should be designed as 
\begin{equation}\label{eq-opt-phase}
	{\bm \omega_m}={\bf a}_{m,i}^{\rm R}.
\end{equation}
{Notably, this phase design is determined only by statistical CSI. As an example, we can calculate the AoA between the UAV$_i$ and the RIS$_m$ from GPS data of the UAV$_i$. In real-world deployment, there inevitably exists a small offset between the UAV’s GPS coordinates and its true location. Considering the relatively large UAV-RIS distance, the position-induced angular deviation is negligible. For instance, when the UAV-RIS distance is 200 meters, a position deviation of 0.1 meters only results in an angle deviation of at most 0.03 degrees. As this deviation falls within the RIS's 3 dB beamwidth, the robustness of the proposed RIS optimization using statistical CSI can be guaranteed.}

\subsection{Reflection Amplitude Design}

Based on the optimal reflection phase design given in \eqref{eq-opt-phase}, $\eta_i$ is determined by the reflection amplitude:
\begin{equation}
	\begin{aligned}
		{\eta _i}&  = \frac{{{N^{\rm{B}}}{E_i}}}{{\left( {\kappa  + 1} \right)\sigma _{\rm{B}}^2}}\sum\nolimits_{k = 1}^3 {\rho _{k,i}^2\rho _k^2\left( {\kappa \left\| {{{\bm{\alpha }}_k}} \right\|_1^2 + {{\left\| {{{\bm{\alpha }}_k}} \right\|}^2}} \right)} \\
		& + \frac{{2\kappa {E_i}\sum\nolimits_{k = 1}^3 {\rho _{k,i}^4\rho _k^4{{\left\| {{{\bm{\alpha }}_k}} \right\|}^2}\left\| {{{\bm{\alpha }}_k}} \right\|_1^2}  + \sum\nolimits_{k = 1}^3 {\rho _{k,i}^4\rho _k^4{{\left\| {{{\bm{\alpha }}_k}} \right\|}^4}} }}{{\left( {\kappa  + 1} \right)\sigma _{\rm{B}}^2\sum\nolimits_{k = 1}^3 {\rho _{k,i}^2\rho _k^2\left( {\kappa \left\| {{{\bm{\alpha }}_k}} \right\|_1^2 + {{\left\| {{{\bm{\alpha }}_k}} \right\|}^2}} \right)} }},
	\end{aligned}
\end{equation}
where ${\bm{\alpha}}_k$ is subject to the power budget and the maximal reflection amplitude. According to the power constraint of RIS$_k$ given in \eqref{eq-RIS-power}, we have 
\begin{equation}
	{\mathbb E}\left\{ {{{\left\| {{{\bf{\Gamma }}_k}\left( {{{\bf{g}}_{k,i}}x + {\bf{z}}_k^{\rm{R}}} \right)} \right\|}^2}} \right\} = \left( {{E_{i}}\rho _{k,i}^2 + \sigma _{\rm{R}}^2} \right){\left\| {{{\bm{\alpha }}_k}} \right\|^2} \le {E_{\rm{R}}}.
\end{equation}
Given that $1 \le | \gamma_{k,n} | \le \gamma_{\max}$, the reflection amplitude ${\bm{\alpha}}_k$ should satisfy 
\begin{equation}
	{\left\| {{{\bm{\alpha }}_k}} \right\|^2} = \min \left\{ {{N^{\rm{R}}}\gamma _{\max }^2,{{{E_{\rm{R}}}}}/{({{E_{i}}\rho _{k,i}^2 + \sigma _{\rm{R}}^2}})} \right\}.
\end{equation}

Using ${{\partial \| {{{\bm{\alpha }}_n}} \|_1^2}}/{{\partial {\alpha _{n,m}}}} = 2{\| {{{\bm{\alpha }}_n}} \|_1}$ and ${{\partial \| {{{\bm{\alpha }}_n}} \|_2^2}}/{{\partial {\alpha _{n,m}}}} = 2{\alpha _{n,m}}$, we derive ${{\partial {\eta _i}}}/{{\partial {\alpha _{n,m}}}}$ as
\begin{equation}
	\frac{{\partial {\eta _i}}}{{\partial {\alpha _{n,m}}}} = \frac{{\rho _{n,i}^2\rho _n^2{E_i}}}{{\left( {\kappa  + 1} \right)\sigma _{\rm{B}}^2}}\frac{{\sum\nolimits_{k = 1}^3 {\rho _{k,i}^2\rho _k^2{Y_{k,n,i}}} }}{{{{\left( {\sum\nolimits_{k = 1}^3 {\rho _{k,i}^2\rho _k^2\left( {\kappa \left\| {{{\bm{\alpha }}_k}} \right\|_1^2 + {{\left\| {{{\bm{\alpha }}_k}} \right\|}^2}} \right)} } \right)}^2}}},
\end{equation}
where $Y_{k,n,i}$ is given in \eqref{eq-Ykni}. Given that ${N^{\rm{B}}}\kappa \| {{{\bm{\alpha }}_k}} \|_1^2 + {N^{\rm{B}}}{\| {{{\bm{\alpha }}_k}} \|^2} \ge 2\kappa \| {{{\bm{\alpha }}_k}} \|_1^2 + {\| {{{\bm{\alpha }}_k}} \|^2}$ and $\sum\nolimits_{t = 1}^3 {\rho _{t,i}^2\rho _t^2( {\kappa \| {{{\bm{\alpha }}_t}} \|_1^2 + {{\| {{{\bm{\alpha }}_t}} \|}^2}} )}  \ge \rho _{k,i}^2\rho _k^2{\| {{{\bm{\alpha }}_k}} \|^2}$, we conclude that $Y_{k,n,i} > 0$, and thus $\eta_i$ is a monotonically increasing function of $\alpha_{n,m}$. Adopting the maximum reflection amplitude to enhance the received signal is a feasible approach. This is because the reflection amplitude significantly boosts the received signal power, while the amplified noise remains negligible compared to the BS-inherent noise.
\begin{figure*}
	\begin{equation}\label{eq-Ykni}
		\begin{aligned}
			{Y_{k,n,i}}&=2\left( {\kappa {{\left\| {{{\bm{\alpha }}_n}} \right\|}_1} + {\alpha _{n,m}}} \right)\left( {{N^{\rm{B}}}\kappa \left\| {{{\bm{\alpha }}_k}} \right\|_1^2 + {N^{\rm{B}}}{{\left\| {{{\bm{\alpha }}_k}} \right\|}^2}} \right)\sum\nolimits_{t = 1}^3 {\rho _{t,i}^2\rho _t^2\left( {\kappa \left\| {{{\bm{\alpha }}_t}} \right\|_1^2 + {{\left\| {{{\bm{\alpha }}_t}} \right\|}^2}} \right)} \\
			&- 2\left( {\kappa {{\left\| {{{\bm{\alpha }}_n}} \right\|}_1} + {\alpha _{n,m}}} \right)\left( {2\kappa \left\| {{{\bf{\alpha }}_k}} \right\|_1^2 + {{\left\| {{{\bm{\alpha }}_k}} \right\|}^2}} \right)\rho _{k,i}^2\rho _k^2{\left\| {{{\bm{\alpha }}_k}} \right\|^2}\\
			&+ \rho _{n,i}^2\rho _n^2\left( {\left( {4\kappa {{\left\| {{{\bm{\alpha }}_n}} \right\|}_1} + 2{\alpha _{n,m}}} \right){{\left\| {{{\bm{\alpha }}_n}} \right\|}^2} + \left( {2\kappa \left\| {{{\bm{\alpha }}_n}} \right\|_1^2 + {{\left\| {{{\bm{\alpha }}_n}} \right\|}^2}} \right)2{\alpha _{n,m}}} \right)\left( {\kappa \left\| {{{\bm{\alpha }}_k}} \right\|_1^2 + {{\left\| {{{\bm{\alpha }}_k}} \right\|}^2}} \right)
		\end{aligned}
	\end{equation}
	%\vspace{-0.5cm}
\hrulefill
\end{figure*}

When the power budget $E_{\rm{R}}$ is sufficiently large, that is, ${E^{\rm{R}}} \ge {N^{\rm{R}}}( {{E_{i}}\rho _{k,i}^2 + \sigma _{\rm{R}}^2} )\gamma _{\max }^2$, the reflection amplitude that maximizes $\eta_i$ is $\alpha_{n,m} = \gamma_{\max}$ for $n \in \{1,2,3\}$ and $m \in \{1,2,\ldots,N^{\rm{R}}\}$. Conversely, when ${E^{\rm{R}}} < {N^{\rm{R}}}( {{E_{i}}\rho _{k,i}^2 + \sigma _{\rm{R}}^2} )\gamma _{\max }^2$, the reflection amplitudes of RIS$_k$ must satisfy 
\begin{equation}
	\left\{ \begin{aligned}
		&\sum\nolimits_{n = 1}^{{N^{\rm{R}}}} {\alpha _{k,n}^2}  \le {{{E_{\rm{R}}}}}{({{E_i}\rho _{k,i}^2 + \sigma _{\rm{R}}^2})}, \\
		&{\alpha _{k,n}} \le {\gamma _{\max }},~n \in \left\{ {1,2, \ldots ,{N^{\rm{R}}}} \right\}.
	\end{aligned} \right.
\end{equation}
In this case, the power budget of RIS$_k$ should be fully utilized. However, not all elements can simultaneously reach the maximum reflection amplitude. For RIS$_k$, the contribution of each reflection to $\eta_i$ is identical. Swapping the values of $\alpha_{k,n}$ and $\alpha_{k,m}$ does not affect $\eta_i$, as $\| {{{\bm{\alpha }}_k}} \|_1^2$ and $\| {{{\bm{\alpha }}_k}} \|_2^2$ remain unchanged. Therefore, it is reasonable to infer that $\eta_i$ is maximized when all elements adopt the same achievable reflection amplitude. In summary, we design the reflection amplitude as 
\begin{equation}\label{eq-opt-amp}
	{\alpha _{k,n}}={\rm{  min}}\left\{ {{\gamma _{\max }},\sqrt {{{{E_{\rm{R}}}}}/{{\left({N^{\rm{R}}} {{E_i}\rho _{k,i}^2 + {N^{\rm{R}}}\sigma _{\rm{R}}^2} \right)}}} } \right\}.
\end{equation}

In a typical low-altitude communication scenario operating at 3.55 GHz ($N^{\rm{R}} = 10 \times 10$, $\sigma_{\rm{R}}^2 = -100$ dBm, $E_i = 20$ dBm, $E_{\rm{R}} = 10$ dBW, and UAV$_i$ is 300 meters away from the RIS), the quantity $\sqrt {\frac{{{E_{\rm{R}}}}}{{{N^{\rm{R}}}( {{E_i}\rho _{k,i}^2 + \sigma _{\rm{R}}^2} )}}} $ is on the order of $10^4$, which is significantly larger than $\gamma_{\max}$. In this case, it is safe to set $\alpha_{k,n} = \gamma_{\max}$ for $n \in \{1,2,3\}$ and $m \in \{1,2,\ldots,N^{\rm{R}}\}$. 

{In summary, the upper bound of noise power ratio ${\beta _{k,i}^{{\rm Upp},2}}$ can be used to assess whether the noise introduced by active RIS can be ignored. When ${\beta _{k,i}^{{\rm Upp},2}}$ is lower than a tolerable threshold of noise ratio, the RIS reflection phase and amplitude that maximize the approximate lower bound of the average SNR are given by \eqref{eq-opt-phase} and \eqref{eq-opt-amp}, {respectively}.}

{\subsection{Performance Evaluation}}
{Using the closed-form expression of the optimized ${\bm{\omega}}_k$ and ${\bm{\alpha}}_k$, the approximate lower bound of receive SNR is then maximized as}
\begin{equation}
	\begin{aligned}
		{\eta _i} = \frac{{{E_i}{N^{\rm{R}}}\gamma _{\max }^2}}{\left( {\kappa  + 1} \right){\sigma _{\rm{B}}^2}}\left( {N^{\rm{B}}}\left( {\kappa {N^{\rm{R}}} + 1} \right)\sum\nolimits_{k = 1}^3 {\rho _{k,i}^2\rho _k^2}  \right.\\
		+\left. \frac{{\left( {2\kappa {N^{\rm{R}}} + 1} \right)\sum\nolimits_{k = 1}^3 {\rho _{k,i}^4\rho _k^4} }}{{\left( {\kappa {N^{\rm{R}}} + 1} \right)\sum\nolimits_{k = 1}^3 {\rho _{k,i}^2\rho _k^2} }} \right),
	\end{aligned}
\end{equation}
{showing} that ${\bf \eta}_i$ grows quadratically with the RIS size $N^{\rm R}$ and the maximum reflection amplitude $\gamma_{\max}$. 

Considering scenarios where the distance from each RIS to the BS is equal ($d=d_k$ yields $\rho_0 = \rho_k$ for $\forall k$) and $\kappa {N^{\rm{R}}} + 1 \approx \kappa {N^{\rm{R}}}$, $\eta_i$  can be simplified as 
\begin{equation}
	{\eta _i} = \frac{{{E_i}\rho _0^2 \left\| {\bf{\Gamma }} \right\|_{\max }^2}}{{\left( {\kappa  + 1} \right)\sigma _{\rm{B}}^2}}\left( \kappa{{N^{\rm{B}}} {N^{\rm{R}}}\sum\nolimits_{k = 1}^3 {\rho _{k,i}^2}  + 2\frac{{\sum\nolimits_{k = 1}^3 {\rho _{k,i}^4} }}{{\sum\nolimits_{k = 1}^3 {\rho _{k,i}^2} }}} \right),
\end{equation}
where $\| {\bf{\Gamma }} \|_{\max }^2=\max \sum\nolimits_{n = 1}^{{N^{\rm{R}}}} {\gamma _{k,i}^2} ={N^{\rm{R}}}\gamma _{\max }^2$ is defined as the maximal array gain of the RIS. For the constrained reflection amplitude, that is, $\gamma _{\max }^2 \le \frac{{\left( {{N^{\rm{B}}}\kappa  + 1} \right){N^{\rm{R}}}}}{{{N^{\rm{B}}} + {N^{\rm{R}}} + \kappa {N^{\rm{R}}}}}$, the upper bound noise power ratio $\beta _{k,i}^{{\rm{Upp}},2}$ can be rewritten as 
\begin{equation}
\beta _{k,i}^{{\rm{Upp}},2}=	\frac{{\rho _0^2\left\| {\bf{\Gamma }} \right\|_{\max }^2}}{{\kappa  + 1}}\left( {2{N^{\rm{B}}}\sqrt {\frac{\kappa }{{{N^{\rm{R}}}}}} {\gamma _{\max }} + 2\left( {{N^{\rm{B}}}\kappa  + 1} \right)} \right).
\end{equation}
The path loss $\rho _0^2$ affects both $\eta_i$ and $\beta _{k,i}^{{\rm{Upp}},2}$ in the same manner. When  $\rho _0^2$ varies with  the distance between the BS and RIS, its impact can be offset by properly choosing $N^{\rm R}$ and $\gamma_{\max}$, that is, adjusting $\| {\bf{\Gamma }} \|_{\max }^2$ to keep ${{\rho _0^2\| {\bf{\Gamma }} \|_{\max }^2}}$ constant. However, $N^{\rm R}$ and $\gamma_{\max}$ have different effects on $\eta_i$ and $\beta _{k,i}^{{\rm{Upp}},2}$. As the RIS deployed farther from the BS, $\rho _0^2$ decreases, requiring larger $N^{\rm R}$ and $\gamma_{\max}$. 

Considering two extreme cases: (i) Increasing only $\gamma_{\max}$, in which case $\eta_i$ remains unchanged, but $\beta _{k,i}^{{\rm{Upp}},2}$ increases. {In this case, the RIS-induced noise becomes increasingly non-negligible, leading to a degradation in the average SNR $\eta^{\rm ave}_{i}$ despite the unchanged lower bound $\eta_i$.} (ii) Increasing only $N^{\rm R}$, in which case $\eta_i$ can be improved and $\beta _{k,i}^{{\rm{Upp}},2}$ decreases, justifying the ignorance of the RIS-introduced noise.  {When ${{\rho _0^2\| {\bf{\Gamma }} \|_{\max }^2}}$ is kept constant, increasing $N^{\rm R}$ and reducing $\gamma_{\max}$ clearly decreases $\beta _{k,i}^{{\rm{Upp}},2}$.} The above analysis shows that, when the RIS array gain is fixed, a larger-sized RIS with a smaller reflection amplitude can achieve better uplink enhancement than a smaller-sized RIS with a larger reflection amplitude. {On the basis of \eqref{eq-beta1}, with appropriate parameter selection to ensure a sufficiently small noise power ratio, active RISs always benefits from its amplification capability and the system performance can be streamlined by omitting the consideration of RIS-induced noise.} 

{In practice, deploying the RIS closer to the BS to achieve a {larger} $\rho _0^2$ can enhance the UAV signals at lower power consumption. However, a larger $\rho _0^2$ also increases the noise power ratio, making the RIS-introduced noise non-negligible. It is worthwhile to identify the minimum BS–RIS distance, below which the amplification gain of active RIS fails to compensate for its introduced noise. To derive useful insights, we assume identical BS-RIS distance $d_k=d$, and let $\eta_{\rm active}={\eta_i^{\rm LB}}$ and $\eta_{\rm passive}=\eta _i^{{\rm{LB}}}( {\sigma _{\rm{R}}^2 = 0,{\gamma _{\max }} = 1} )$ denote the lower bound SNR of active-RIS and passive-RIS systems, respectively. Calculations derived from Appendix \ref{APP-last} reveal that $\eta_{\rm active}<\eta_{\rm passive}$ holds once the BS–RIS distance satisfies
\begin{equation}\label{eq:d_min}
	d < \sqrt {\frac{{{C_2}\sigma _{\rm{R}}^2}}{{{C_3}}}\frac{{\gamma _{\max }^2}}{{\gamma _{\max }^2 - 1}}} \buildrel \Delta \over = d_{\rm min}, 
	\end{equation}
where $C_3 =  {\sigma _{\rm{B}}^2\left( {\kappa {N^{\rm{R}}} + 1} \right)}$ and 
\begin{equation}
	\begin{aligned}
		C_2  =& \left( {{N^{\rm{B}}}\left( {{\kappa ^2}{N^{{\rm{R}},2}} + 2\kappa {N^{\rm{R}}} + 1} \right) + {N^{\rm{R}}}\left( {\kappa {N^{\rm{R}}} + \kappa  + 1} \right)} \right)\\
		&\times \frac{{{\lambda ^2}}}{{16{\pi ^2}\left( {\kappa  + 1} \right)}}.
	\end{aligned}
\end{equation}
In the extreme case where the BS--RIS distance is smaller than $d_{\rm min}$, the negative impact of noise inherent to active RIS exceeds the benefits of signal amplification, and thus the active RIS system suffers from inferior SNR performance relative to passive RIS. In practical RIS deployment, where $\kappa>1$, $N_{\rm R}>>1$, and $N_{\rm B}>>1$, $d_{\rm min}$ can be approximated by
\begin{equation}
	d_{\rm min}\approx\frac{{\lambda {\sigma _{\rm{R}}}}}{{4\pi {\sigma _{\rm{B}}}}}\sqrt {\frac{{{N^{\rm{R}}}{N^{\rm{B}}}\kappa }}{{\kappa  + 1}}\frac{{\gamma _{\max }^2}}{{\gamma _{\max }^2 - 1}}}. 
\end{equation}
This analytical expression indicates that $d_{\rm min}$ depends primarily on $N_{\rm B}$ and $N_{\rm R}$, with ${\frac{{\gamma _{\max }^2}}{{\gamma _{\max }^2 - 1}}}\approx 1$. Larger arrays require a longer separation between the active RIS and the receiver to outperform the passive RIS in terms of SNR gain.
}

%%%%%%%%%%%%%%%%%%%%%%%%%%%%%%%%%%%%%%%%%%%%%%%%%%%%%%%%%%%%%%%%%%%%%%%%%%%%%%%%%%%%%%%%%%%%%%

\section{Numerical Results}\label{sec:5}
\begin{figure}
	\centering
	\includegraphics[width=0.5\textwidth]{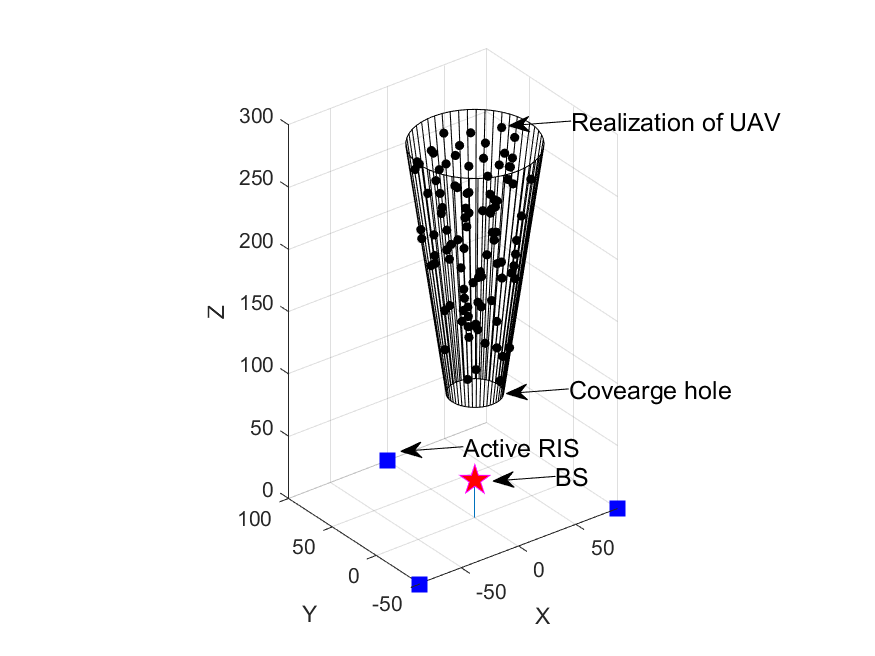}
	\caption{Active RISs-assisted low-altitude communications.}
	\label{FigSys}
	%\vspace{-0.5cm}
\end{figure}

In this section, we present numerical simulations to validate the noise analysis results and to demonstrate the effectiveness of the optimized RIS reflection. {The noise power is given by $\sigma^2=-174+NF+10\log(BW)$, where $NF=10$ is the noise figure and $BW$ is the bandwidth \cite{NoisePower}. In a system operating at $3.55$ GHz with a $2.5$ MHz bandwidth, the noise powers $\sigma_{\rm{B}}^2$ and $\sigma_{\rm{R}}^2$ are fixed at $-100$ dBm.} Unless otherwise specified, the simulation parameters are set as follows. As shown in Fig. \ref{FigSys}, the Cartesian coordinates for the BS and the three active RISs are ${{\bf{p}}^{\rm{B}}} = {[ {0,0,30} ]^{\rm T}}$, ${\bf{p}}_1^{\rm{R}} = {[ {0,100,0} ]^{\rm T}}$, ${\bf{p}}_2^{\rm{R}} = {[ {-50,50\sqrt{3},0} ]^{\rm T}}$, and ${\bf{p}}_3^{\rm{R}} = {[ {50,50\sqrt{3},0} ]^{\rm T}}$, respectively. That is, the three active RISs are deployed at the LoS centers of the three BS sectors, each with a horizontal distance of $d_{\rm BS-RIS} = 100$ meters from the BS. Given the strong LoS link in the BS-RIS channel, the Rician factor $\kappa$ is set to $10$ dB. The downtilt angle of each BS antenna array is $\varpi^{\rm B} = 8^\circ$, leading to a frustum-shaped coverage hole centered at $[0,0,100]^{\rm T}$ and $[0,0,300]^{\rm T}$. In each BS sector, the UPA consists of $N^{\rm B} = 12 \times 8$ antennas, which is a typical configuration for a $3.55$ GHz active antenna unit array. By default, each RIS comprises $N^{\rm R} = 10 \times 10$ elements, with a maximum reflection amplitude $\gamma_{\max} = 10$ for each element. The transmit power of the UAV is $10$ dBm. The numerical results are averaged over 2,500 realizations of the UAV's random position within the coverage hole. The free space path loss can be calculated based on the relative positions of the BS, RISs, and the UAV. The performance of passive RISs is provided as a baseline.

\subsection{Noise Power Ratio}

In this subsection, we verify the analytical results and insights regarding the BS-inherent and RIS-introduced noise obtained in Sec. \ref{sec:3}. Specifically, based on \eqref{eq-beta2} in Theorem \ref{Theo-betaU2}, we focus on the impacts of the reflection amplitude, the size of the RIS, the Rician factor, and the BS--RIS distance on the noise power ratio.

\begin{figure}
	\centering
	\includegraphics[width=0.5\textwidth]{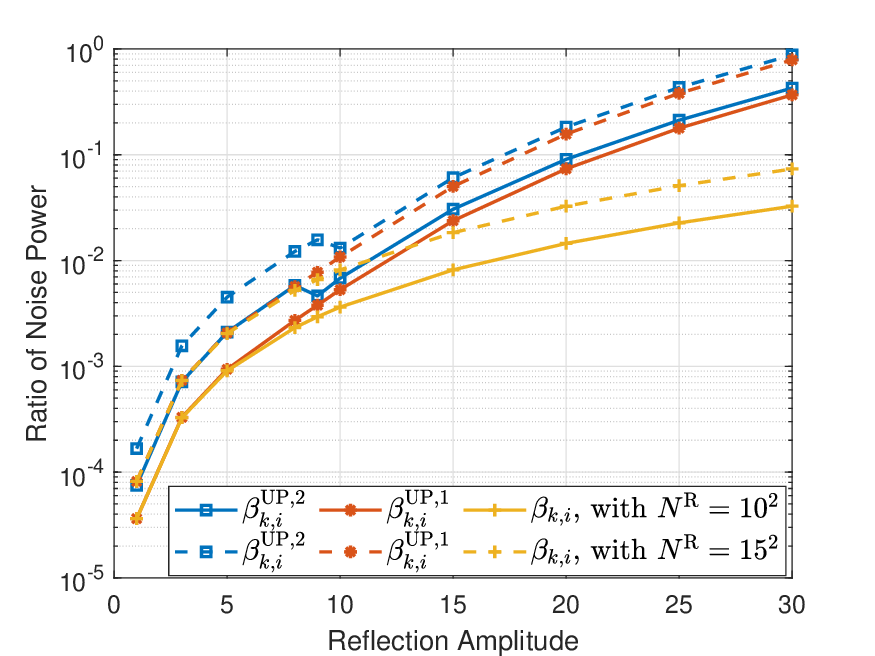}
	\caption{Impacts of $\gamma_{\max }$ and $N^{\rm R}$ on the noise power ratio.}
	\label{FigC-1}
	%\vspace{-0.5cm}
\end{figure}

In Fig. \ref{FigC-1}, we depict how the ratio of {RIS-introduced noise power to BS-inherent noise power} evolves as the maximum reflection amplitude $\gamma_{\max}$ and the RIS size $N^{\rm R}$ increase. The solid and dashed lines correspond to results for different RIS sizes. It is evident that for both types of lines, $\beta_{k,i}$ and $\beta_{k,i}^{\rm UP,1}$ nearly coincide when $\gamma_{\max}$ is small. This alignment occurs because $\beta_{k,i}^{\rm UP,1}$ is an upper bound derived by scaling $\beta_{k,i}$ based on the reflection amplitude constraint $1 \leq |\gamma_{k,n}| \leq \gamma_{\max}$. As $\gamma_{\max}$ approaches 1, narrowing the range of $|\gamma_{k,n}|$, $\beta_{k,i}^{\rm UP,1}$ converges to $\beta_{k,i}$. Conversely, as the reflection amplitude increases, the gap between $\beta_{k,i}$ and $\beta_{k,i}^{\rm UP,1}$ widens. {Considering that the RIS-induced noise increases with the reflection amplitude, $\beta_{k,i}^{\rm UP,1}$, as a loose upper bound for $\beta_{k,i}$, offers a conservative criterion for determining whether the RIS-induced noise can be neglected. Notably, in light of the variation in the actual noise ratio $\beta_{k,i}$, even a large-scale RIS ($N^{\rm R}=15^2$) with a high reflection amplitude ($\gamma_{\max}=30$), the noise it introduces remains less than 0.1 times the BS-inherent noise ($\beta_{k,i}<0.1$).} Although $\beta_{k,i}^{\rm UP,2}$ is looser than $\beta_{k,i}^{\rm UP,1}$, it exhibits a similar growth trend. {Its concise form not only captures the trend of $\beta_{k,i}^{\rm UP,1}$ but also yields valuable insights into the impacts of relevant parameters, as analyzed in Theorem \ref{Theo-betaU2}. To derive these insights, a more relaxed scaling is adopted for $\gamma _{\max }^2 \le{{\left( {{N^{\rm{B}}}\kappa  + 1} \right){N^{\rm{R}}}}}/{({{N^{\rm{B}}} + {N^{\rm{R}}} + \kappa {N^{\rm{R}}}})}$, which results in the non-smooth behavior of $\beta_{k,i}^{\rm UP,2}$, as discussed in Appendix \ref{APP-T2}.}

\begin{figure}
	\centering
	\includegraphics[width=0.5\textwidth]{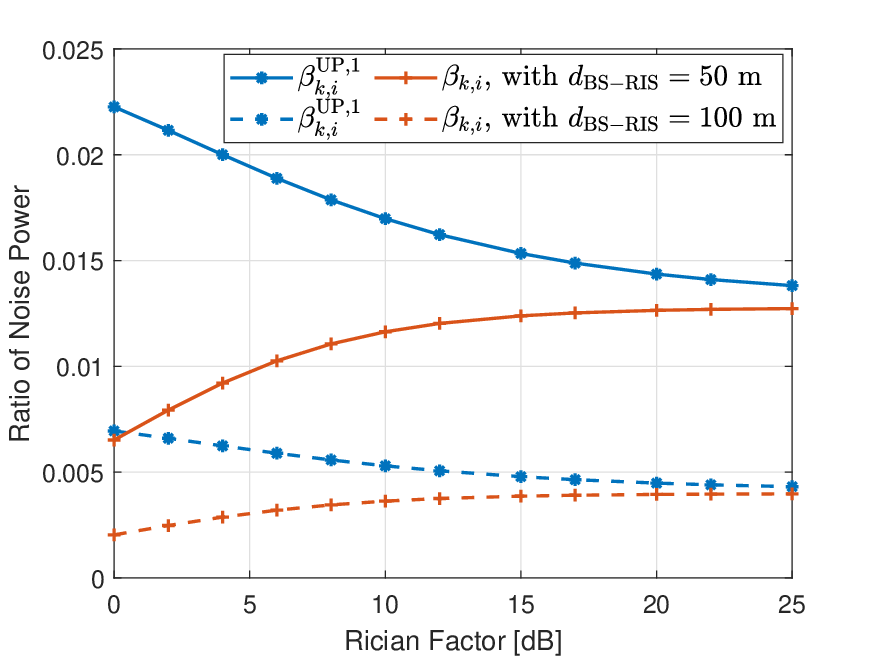}
	\caption{Impacts of $\kappa$ and $\rho_k$ on the noise power ratio.}
	\label{FigC-2}
	%\vspace{-0.5cm}
\end{figure}

Based on Theorem \ref{Theo-betaU2}, we present in Fig. \ref{FigC-2} the impact of path loss and the Rician factor, which are both statistical channel parameters, on the noise power ratio. Since path loss is a function of the distance between the BS and the RIS, denoted as $d_{\rm BS-RIS}$, we present the simulation results in terms of $d_{\rm BS-RIS}$ rather than specific values of $\rho_k$. The solid and dashed lines represent the results for different distances. As the RIS is deployed farther from the BS, the RIS-introduced noise experiences more severe path loss. Consequently, compared to the constant BS-inherent noise, the RIS-introduced noise becomes increasingly negligible. This trend aligns with the analysis in Theorem \ref{Theo-betaU2}. Although the closed-form expressions do not explicitly reveal the impact of the Rician factor on the noise power ratio, Fig. \ref{FigC-2} shows that $\beta_{k,i}$ exhibits a slow increase and eventually stabilizes as $\kappa$ increases. This behavior is attributed to the channel gradually hardening into the LoS path, with the RIS-introduced noise reaching the BS only through the statistically invariant LoS component. Moreover, the convergence of $\beta_{k,i}$ and $\beta_{k,i}^{\rm UP,1}$ as $\kappa$ increases indicates that the upper bound $\beta_{k,i}^{\rm UP,1}$ can better represent the true noise power ratio in strong LoS scenarios.

\subsection{SNR Gain with the Optimized RIS Reflection}

\begin{figure}
	\centering
	\includegraphics[width=0.5\textwidth]{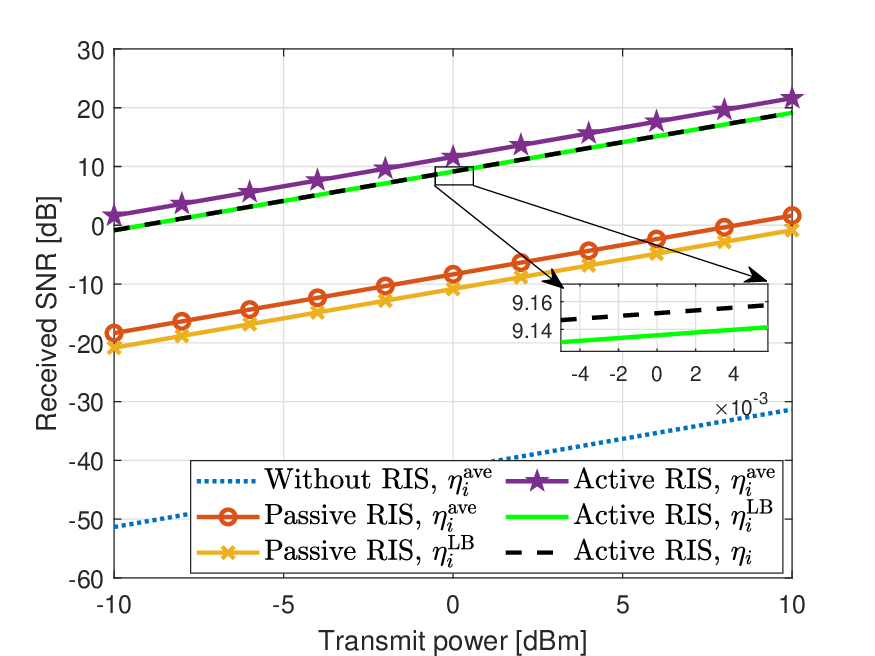}
	\caption{Comparison of the receive SNR in active and passive RIS-assisted systems. }
	\label{FigA-1}
	%\vspace{-0.5cm}
\end{figure}

This subsection evaluates the effectiveness of the optimized RIS reflection. Fig. \ref{FigA-1} compares the received SNR of the UAV signal at the BS in three scenarios: without RIS, with passive RIS, and with active RIS. When RISs are not configured, they essentially act as metallic surfaces that only reflect electromagnetic waves specularly. This condition is equivalent to having no RIS assistance. Due to the downtilt configuration of the BS antenna array, the receive SNR $\eta_i^{\rm ave}$ remains significantly low even under tri-sector reception, in the absence of RISs.
Introducing RISs, whether passive or active, can substantially improve the receive SNR. Notably, the active RIS with $\gamma_{\max}=10$ provides an SNR improvement of approximately 20 dB compared to the passive RIS. Although there is an SNR gap of about 2.5 dB between the average SNR $\eta_i^{\rm ave}$ and its lower bound $\eta_i^{\rm LB}$ in both cases, their growth trends remain consistent. This consistency validates the use of $\eta_i^{\rm LB}$ as a reliable indicator for assessing RIS impact on SNR. Furthermore, the curve of $\eta_i$, which neglects RIS-introduced noise, nearly overlaps with that of $\eta_i^{\rm LB}$, confirming that the noise introduced by the RIS amplification circuit is negligible after undergoing the RIS--BS path loss.

\begin{figure}
	\centering
	\includegraphics[width=0.5\textwidth]{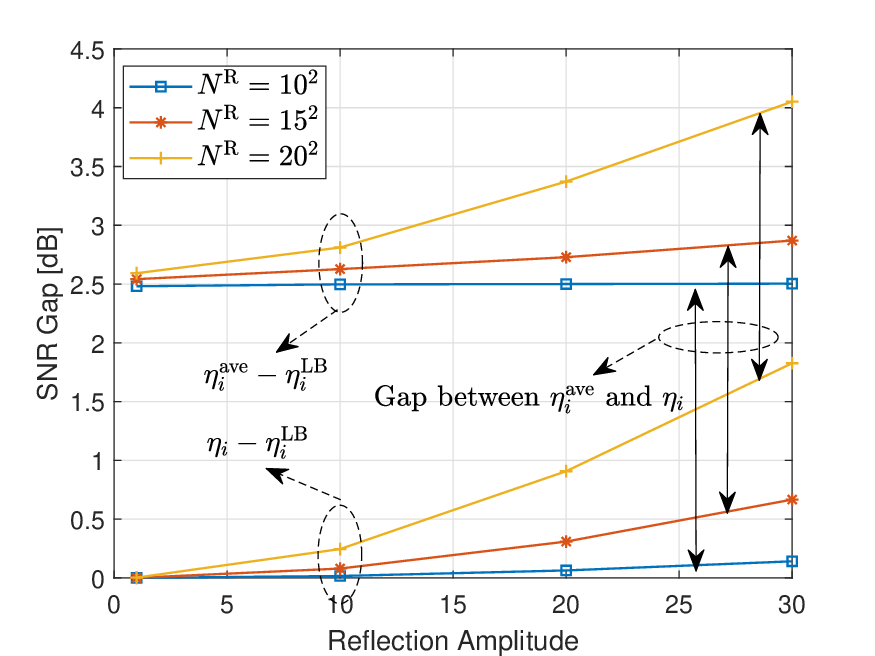}
	\caption{Gaps between the average and lower bounds of the receive SNR.}
	\label{FigA-2}
	%\vspace{-0.5cm}
\end{figure}

Fig. \ref{FigA-2} illustrates the tightness of the derived SNR lower bounds under varying values of $\gamma_{\max}$ and $N^{\rm R}$ by showing the gaps between the average and lower-bound SNRs. As the maximum reflection amplitude $\gamma_{\max}$ increases, the gap $\eta_i^{\rm ave} - \eta_i^{\rm LB}$ becomes larger, indicating a reduced tightness of the lower bound due to the inequality ${\mathbb E}\{X\}/{\mathbb E}\{Y\} \le {\mathbb E}\{X/Y\}$. Fortunately, this performance loss diminishes for smaller RIS sizes.
On the other hand, $\eta_i$ exceeds $\eta_i^{\rm LB}$ as it neglects the RIS-introduced noise. Hence, the gap $\eta_i - \eta_i^{\rm LB}$ also increases with RIS-introduced noise power, which is affected by higher $\gamma_{\max}$ or larger $N^{\rm R}$. Together, these opposing trends yield a relatively stable gap between $\eta_i^{\rm ave}$ and $\eta_i$, reinforcing that $\eta_i$ serves as a robust lower bound for $\eta_i^{\rm ave}$ across different parameter settings.

\begin{figure}
	\centering
	\includegraphics[width=0.5\textwidth]{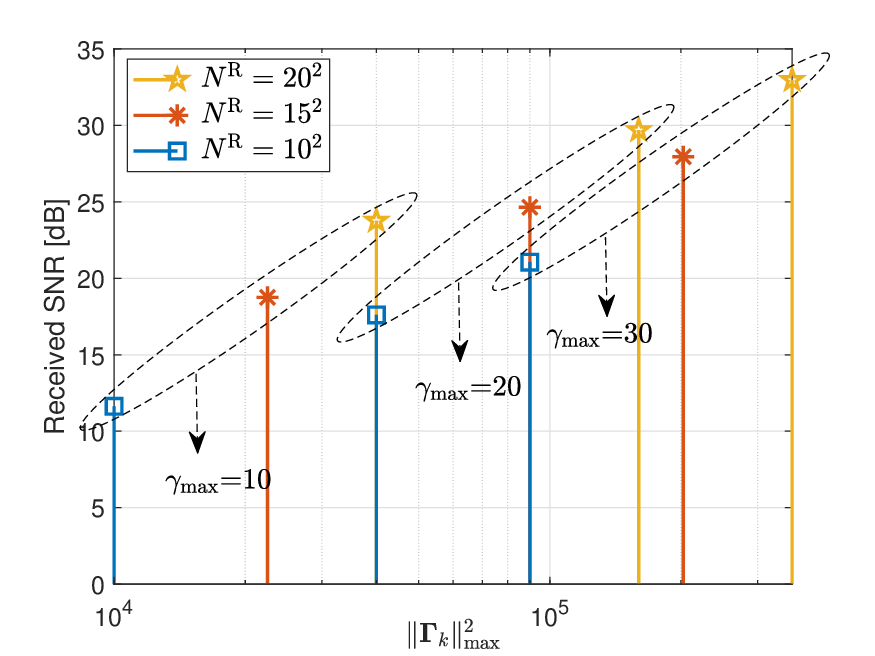}
	\caption{Receive SNR versus the array gain of active RISs.}
	\label{FigA-3}
	%\vspace{-0.5cm}
\end{figure}

Increasing the number of RIS elements $N^{\rm R}$ and the maximum reflection amplitude $\gamma_{\max}$ are two effective strategies to enhance the array gain $\|{\bf{\Gamma}}_k\|^2_{\rm max}=N^{\rm R}\gamma^2_{\max}$ of an active RIS, each affecting the received SNR in distinct ways. Fig. \ref{FigA-3} illustrates how the average received SNR $\eta_i^{\rm ave}$ varies with $\|{\bf{\Gamma}}_k\|^2_{\rm max}$, offering insights into the trade-off between $N^{\rm R}$ and $\gamma_{\max}$ under a fixed array gain.
When the array gain is fixed, for example, at $\|{\bf{\Gamma}}_k\|^2_{\rm max} = 4 \times 10^4$, an active RIS with $N^{\rm R} = 20^2$ and $\gamma_{\max} = 10$ achieves higher received SNR than one with $N^{\rm R} = 10^2$ and $\gamma_{\max} = 20$. This result indicates that increasing RIS size yields greater SNR benefits than merely enhancing amplification. This observation is intuitive: while increasing $\gamma_{\max}$ boosts the signal, it also raises the level of RIS-introduced noise, which negatively affects the SNR.
This advantage of increasing RIS size is further supported by comparing an active RIS with $N^{\rm R} = 20^2$ and $\gamma_{\max} = 20$ to another with $N^{\rm R} = 15^2$ and $\gamma_{\max} = 30$. Despite a lower array gain, the former achieves a higher SNR due to its larger surface. These findings suggest that, when enhancing UAV uplink transmission using RISs, prioritizing a larger RIS size is more advantageous than maximizing the reflection amplitude.

\section{Conclusion}\label{sec:6}
This study proposed a low-cost solution for enhancing uplink transmission in low-altitude communications within existing infrastructure. While commercial BSs typically employ downtilt antennas to optimize terrestrial coverage, this conventional setup inadvertently creates coverage holes in low-altitude airspace, particularly compromising uplink services for UAVs requiring high-definition video upload. Notably, deploying dedicated uptilt BSs for aerial coverage involves prohibitive infrastructure costs and introduces additional interference to existing networks. 

To overcome these challenges, we introduced active RISs into legacy networks as a cost-effective coverage enhancement solution. A JS-MSC scheme was proposed to harness spatial diversity through tri-sector BS cooperation. Based on this receive scheme, we established an upper bound for the noise power ratio between active RIS-introduced and BS-inherent components. This analysis revealed that, under typical commercial network parameters, RIS-introduced noise constitutes a negligible portion (less than $3\%$) of the total noise power. This key finding enabled the derivation of an approximate lower bound for the received SNR.
To improve the SNR, we formulated and solved an optimization problem for designing the reflection coefficients under active RIS power constraints. Numerical results validated both the tightness of the noise ratio bound under low reflection amplitudes and the significant SNR improvement achieved through optimized active RIS design. The proposed active RIS-assisted system demonstrated an SNR enhancement of approximately 20 dB with a maximum 10 dB reflection amplitude, substantially outperforming conventional passive RIS solutions.

% or


\appendices
\section{}\label{APP-L1}
By expanding ${\mathbb E}\{ {{{\| {{{\bf{h}}_i}} \|}^2}} \}$, we have
\begin{equation}\label{APP-L1-1}
	\begin{aligned}
		{\mathbb E}\left\{ {{{\left\| {{{\bf{h}}_i}} \right\|}^2}} \right\} = \sum\nolimits_{k = 1}^3 {\mathbb E}{\left\{ {{\bf{g}}_{k,i}^{\rm{H}}{\bf{\Gamma }}_k^{\rm{H}}{\bf{G}}_k^{\rm{H}}{{\bf{G}}_k}{{\bf{\Gamma }}_k}{{\bf{g}}_{k,i}}} \right\}}. 
	\end{aligned}
\end{equation}
As ${{{\bf{g}}_{k,i}}}$ and ${{{\bm{\Gamma }}_k}}$ are constant,  \eqref{APP-L1-1} can be simplified as 
\begin{equation}\label{App-L1-sum3}
	{\mathbb E}\left\{ {{{\left\| {{{\bf{h}}_i}} \right\|}^2}} \right\} = \sum\nolimits_{k = 1}^3 {{\bf{g}}_{k,i}^{\rm{H}}{\bf{\Gamma }}_k^{\rm{H}}{\mathbb E}\left\{ {{\bf{G}}_k^{\rm{H}}{{\bf{G}}_k}} \right\}{{\bf{\Gamma }}_k}{{\bf{g}}_{k,i}}} .
\end{equation}
Separating statistical LoS and fast-varying NLoS components of the BS--RIS$_k$ channel contains  yields 
\begin{equation}\label{APP-L1-NLoS}
	\begin{aligned}
		{\mathbb E}\left\{ {{\bf{G}}_k^{\rm{H}}{{\bf{G}}_k}} \right\}
		 = {\bf{G}}_k^{{\rm{L,H}}}{\bf{G}}_k^{\rm{L}} + {\mathbb E}\left\{ {{\bf{G}}_k^{{\rm{N,H}}}{\bf{G}}_k^{\rm{N}}} \right\}.
	\end{aligned}
\end{equation}
As the Rayleigh channel is adopted for ${\bf{G}}_k^{\rm{N}}$, we have 
\begin{equation}\label{eq-EGHG}
	{\mathbb E}\left\{ {{\bf{G}}_k^{{\rm{N,H}}}{\bf{G}}_k^{\rm{N}}} \right\}=\frac{{{N^{\rm{B}}}\rho _k^2}}{{\kappa  + 1}}{{\bf{I}}_{ {N^{\rm{R}}}}}.
\end{equation}
Substituting \eqref{APP-L1-NLoS} and expressions of LoS channel components into \eqref{App-L1-sum3}, its $k$-th item can be expressed as
\begin{equation}\label{APP-L1-end}
	\begin{aligned}
		&{\bf{g}}_{k,i}^{\rm{H}}{\bf{\Gamma }}_k^{\rm{H}}{\mathbb E}\left\{ {{\bf{G}}_k^{\rm{H}}{{\bf{G}}_k}} \right\}{{\bf{\Gamma }}_k}{{\bf{g}}_{k,i}}\\
		& = \frac{{{N^{\rm{B}}}\rho _k^2}}{{\kappa  + 1}}{\bf{g}}_{k,i}^{\rm{H}}{\bf{\Gamma }}_k^{\rm{H}}\left( {\kappa {\bf{a}}_{k}^{{\rm{RD}}}{\bf{a}}_{k}^{{\rm{RD}},{\rm{H}}} + {{\bf{I}}_{ {N^{\rm{R}}}}}} \right){{\bf{\Gamma }}_k}{{\bf{g}}_{k,i}}\\
		&= \frac{{{N^{\rm{B}}}\rho _k^2\rho _{k,i}^2}}{{\kappa  + 1}}\left( {\kappa {{\left| {{\bm{\gamma }}_k^{\rm{H}}{\bf{a}}_{k,i}^{\rm{R}}} \right|}^2} + {{\left\| {{{\bm{\gamma }}_k}} \right\|}^2}} \right),
	\end{aligned}
\end{equation}
where ${\bf{a}}_{k,i}^{\rm{R}} = {\bf{a}}_{k,i}^{{\rm{RA,*}}} \odot {\bf{a}}_{k}^{{\rm{RD}}}$. The proof is completed by bringing \eqref{APP-L1-end} back to \eqref{App-L1-sum3}.

\section{}\label{APP-L2}
By decoupling the first two ${{\bf{G}}_k}$ in ${\mathbb E}\{ {{\bf{G}}_k^{\rm{H}}{{\bf{G}}_k}{{\bf{\Gamma }}_k}{\bf{\Gamma }}_k^{\rm{H}}{\bf{G}}_k^{\rm{H}}{{\bf{G}}_k}} \}$ as $ {\bf{G}}_k^{\rm{L}}+ {\bf{G}}_k^{\rm{N}}$,  we have
\begin{equation}
	{\mathbb E}\left\{ {{\bf{G}}_k^{\rm{H}}{{\bf{G}}_k}{{\bf{\Gamma }}_k}{\bf{\Gamma }}_k^{\rm{H}}{\bf{G}}_k^{\rm{H}}{{\bf{G}}_k}} \right\} = {T_1} + {T_2} + {T_3} + {T_4},
\end{equation}
where
\begin{subequations}
	\begin{align}
    {T_1}={\mathbb E}\left\{ {{\bf{G}}_k^{{\rm{L,H}}}{\bf{G}}_k^{\rm{L}}{{\left| {{{\bf{\Gamma }}_k}} \right|}^2}{\bf{G}}_k^{\rm{H}}{{\bf{G}}_k}} \right\},\\
     {T_2}=	{\mathbb E}	\left\{ {{\bf{G}}_k^{{\rm{L,H}}}{\bf{G}}_k^{\rm{N}}{{\left| {{{\bf{\Gamma }}_k}} \right|}^2}{\bf{G}}_k^{\rm{H}}{{\bf{G}}_k}} \right\},\\
	{T_3}=	{\mathbb E}\left\{ {{\bf{G}}_k^{{\rm{N,H}}}{\bf{G}}_k^{\rm{L}}{{\left| {{{\bf{\Gamma }}_k}} \right|}^2}{\bf{G}}_k^{\rm{H}}{{\bf{G}}_k}} \right\},\\
	{T_4}={\mathbb E}	\left\{ {{\bf{G}}_k^{{\rm{N,H}}}{\bf{G}}_k^{\rm{N}}{{\left| {{{\bf{\Gamma }}_k}} \right|}^2}{\bf{G}}_k^{\rm{H}}{{\bf{G}}_k}} \right\}.
	\end{align} 
\end{subequations}
{are} calculated in the following.
\subsection{Closed-form expression for $T_1$}
Substituting ${{\bf{G}}_k}= {\bf{G}}_k^{\rm{L}}+ {\bf{G}}_k^{\rm{N}}$ into $T_1$, we have ${T_1} = {T_{11}} + {T_{12}} + {T_{13}} + {T_{14}} $, where
\begin{subequations}
	 \begin{align}
		{T_{11}}&={\bf{G}}_k^{{\rm{L,H}}}{\bf{G}}_k^{\rm{L}}{\left| {{{\bf{\Gamma }}_k}} \right|^2}{\bf{G}}_k^{{\rm{L,H}}}{\bf{G}}_k^{\rm{L}},\\
		{T_{12}}&={\bf{G}}_k^{{\rm{L,H}}}{\bf{G}}_k^{\rm{L}}{\left| {{{\bf{\Gamma }}_k}} \right|^2}{\bf{G}}_k^{{\rm{L,H}}}{\mathbb E}\left\{ {{\bf{G}}_k^{\rm{N}}} \right\},\\
		{T_{13}}&={\bf{G}}_k^{{\rm{L,H}}}{\bf{G}}_k^{\rm{L}}{\left| {{{\bf{\Gamma }}_k}} \right|^2}{\mathbb E}\left\{ {{\bf{G}}_k^{{\rm{N,H}}}} \right\}{\bf{G}}_k^{\rm{L}},\\
		{T_{14}}&={\bf{G}}_k^{{\rm{L,H}}}{\bf{G}}_k^{\rm{L}}{\left| {{{\bf{\Gamma }}_k}} \right|^2}{\mathbb E}\left\{ {{\bf{G}}_k^{{\rm{N,H}}}{\bf{G}}_k^{\rm{N}}} \right\}.
	\end{align}
\end{subequations}
Considering that the average of the small-scale gain of NLoS paths is 0, ${\mathbb E}\{ {{\bf{G}}_k^{\rm{N}}} \}={\mathbb E}\{ {{\bf{G}}_k^{\rm{N,H}}} \}={\bf 0}$ holds, resulting in $T_{12}=T_{13}={\bf 0}$. Substituting \eqref{eq-EGHG} into $T_{14}$, $T_1$ can be given by 
\begin{equation}
	{T_1}{\rm{ = }}{\bf{G}}_k^{{\rm{L,H}}}{\bf{G}}_k^{\rm{L}}{\left| {{{\bf{\Gamma }}_k}} \right|^2}{\bf{G}}_k^{{\rm{L,H}}}{\bf{G}}_k^{\rm{L}} + \frac{{{N^{\rm{B}}}\rho _k^2}}{{\kappa  + 1}}{\bf{G}}_k^{{\rm{L,H}}}{\bf{G}}_k^{\rm{L}}{\left| {{{\bf{\Gamma }}_k}} \right|^2}.
\end{equation}
\subsection{Closed-form expression for $T_2$}
Similarly, $T_2$ can be unfolding into ${T_{21}} + {T_{22}} + {T_{23}} + {T_{24}} $, where
\begin{subequations}
	 \begin{align}
		{T_{21}}&={\bf{G}}_k^{{\rm{L,H}}}{\mathbb E}\left\{ {\bf{G}}_k^{\rm{N}} \right\}{{\left| {{{\bf{\Gamma }}_k}} \right|}^2}{\bf{G}}_k^{{\rm{L,H}}}{\bf{G}}_k^{\rm{L}},\\
		{T_{22}}&={\bf{G}}_k^{{\rm{L,H}}}{\mathbb E}\left\{ {{\bf{G}}_k^{\rm{N}}{{\left| {{{\bf{\Gamma }}_k}} \right|}^2}{\bf{G}}_k^{{\rm{L,H}}}{\bf{G}}_k^{\rm{N}}} \right\},\\
		{T_{23}}&={\bf{G}}_k^{{\rm{L,H}}}{\mathbb E}\left\{ {{\bf{G}}_k^{\rm{N}}{{\left| {{{\bf{\Gamma }}_k}} \right|}^2}{\bf{G}}_k^{{\rm{N,H}}}} \right\}{\bf{G}}_k^{\rm{L}},\\
		{T_{24}}&={\bf{G}}_k^{{\rm{L,H}}}{\mathbb E}\left\{ {{\bf{G}}_k^{\rm{N}}{{\left| {{{\bf{\Gamma }}_k}} \right|}^2}{\bf{G}}_k^{{\rm{N,H}}}{\bf{G}}_k^{\rm{N}}} \right\}.
	\end{align} 
\end{subequations}

Given that the elements of ${\bf{G}}_k^{\rm{N}}$ have a mean of 0, the terms ${\mathbb E}\{ {{\bf{G}}_k^{\rm{N}}} \}$ in $T_{21}$ and ${\mathbb E}\{ {{\bf{G}}_k^{\rm{N}}{{| {{{\bf{\Gamma }}_k}} |}^2}{\bf{G}}_k^{{\rm{N,H}}}{\bf{G}}_k^{\rm{N}}} \}$ in $T_{24}$, which involve the first- and third-order moments of ${\bf{G}}_k^{\rm{N}}$, respectively, are both equal to ${\bf 0}$. 

\begin{lemma}\label{ABC}
	For  matrix ${\bf A}\in {\mathbb C}^{M\times L}$, ${\bf B}\in {\mathbb C}^{L\times K}$, and ${\bf C}\in {\mathbb C}^{K\times N}$ with the element in the $i$-th row and $j$-th column denoted by $a_{i,j}$, $b_{i,j}$, and $c_{i,j}$, respectively, the element in the $m$-th row and $n$-th column of ${\bf ABC}$ can be expressed as
	\begin{equation}
		\left[{\bf ABC}\right]_{m,n} = \sum\nolimits_{j = 1}^K {\sum\nolimits_{i = 1}^L {{a_{m,i}}{b_{i,j}}{c_{j,n}}} } .
	\end{equation}
\end{lemma}

Utilizing Lemma \ref{ABC} with ${\bf A}={{\bf{G}}_k^{{\rm{N,H}}}}$, ${\bf B}={{{| {{{\bf{\Gamma }}_k}} |}^2}{\bf{G}}_k^{{\rm{L,H}}}}$, and ${\bf C} = {{\bf{G}}_k^{{\rm{N}}}}$ for $T_{22}$, we obtain
\begin{equation}\label{eq-T22-Emn}
	\begin{aligned}
		&{\left[ {\mathbb E}{\left\{ {{\bf{G}}_k^{\rm{N}}{{\left| {{{\bf{\Gamma }}_k}} \right|}^2}{\bf{G}}_k^{{\rm{L,H}}}{\bf{G}}_k^{\rm{N}}} \right\}} \right]_{m,n}} \\
		= &{\mathbb E}\left\{ {\sum\nolimits_{j = 1}^{N^{\rm B}} {\sum\nolimits_{i = 1}^{N^{\rm R}} {{g_{m,i}}{{\left[ {{{\left| {{{\bf{\Gamma }}_k}} \right|}^2}{\bf{G}}_k^{{\rm{L,H}}}} \right]}_{i,j}}{g_{j,n}}} } } \right\}\\
		 =& \sum\nolimits_{j = 1}^{N^{\rm B}} {\sum\nolimits_{i = 1}^{N^{\rm R}} {{{\left[ {{{\left| {{{\bf{\Gamma }}_k}} \right|}^2}{\bf{G}}_k^{{\rm{L,H}}}} \right]}_{i,j}}{\mathbb E}\left\{ {{g_{m,i}}{g_{j,n}}} \right\}} } ,
	\end{aligned}
\end{equation}
where $g_{m,i}$ is the element in the $m$-th row and $i$-th column of ${\bf G}_k^{\rm N}$. As ${g_{m,i}}$ and ${g_{j,n}}$ are independent when $i\ne n$ and $j\ne m$, \eqref{eq-T22-Emn} can be rewritten as
\begin{equation}
		\begin{aligned}
			{\left[ {\mathbb E}{\left\{ {{\bf{G}}_k^{\rm{N}}{{\left| {{{\bf{\Gamma }}_k}} \right|}^2}{\bf{G}}_k^{{\rm{L,H}}}{\bf{G}}_k^{\rm{N}}} \right\}} \right]_{m,n}} 
			= {{{\left[ {{{\left| {{{\bf{\Gamma }}_k}} \right|}^2}{\bf{G}}_k^{{\rm{L,H}}}} \right]}_{n,m}}{\mathbb E}\left\{ {{g_{m,n}}{g_{m,n}}} \right\}}  .
		\end{aligned}
\end{equation}
For Gaussian variable $g_{m,n}$, ${\mathbb E}\{ {{g_{m,n}}{g_{m,n}}} \}=0$ holds, indicating that $T_{22}={\bf 0}$.

In a similar way, we have
\begin{equation}\label{eq-T23-Emn}
	\begin{aligned}
		&{\left[ {\mathbb E}{\left\{ {{\bf{G}}_k^{\rm{N}}{{\left| {{{\bf{\Gamma }}_k}} \right|}^2}{\bf{G}}_k^{\rm{N,H}}} \right\}} \right]_{m,n}} \\
		=& \sum\nolimits_{j = 1}^{N^{\rm B}} {\sum\nolimits_{i = 1}^{N^{\rm R}} {{{\left[ {{{\left| {{{\bf{\Gamma }}_k}} \right|}^2} } \right]}_{i,j}}{\mathbb E}\left\{ {{g_{m,i}}{g_{n,j}^{*}}} \right\}} } ,
	\end{aligned}
\end{equation}
for $T_{23}$. Based on the independence of ${g_{m,i}}$ and ${g_{j,n}}$, we have ${[ {\mathbb E}{\{ {{\bf{G}}_k^{\rm{N}}{{| {{{\bf{\Gamma }}_k}} |}^2}{\bf{G}}_k^{\rm{N,H}}} \}} ]_{m,n}} = 0$ for $n\ne m$ and 
\begin{equation}
	\begin{aligned}
		&{\left[ {\mathbb E}{\left\{ {{\bf{G}}_k^{\rm{N}}{{\left| {{{\bf{\Gamma }}_k}} \right|}^2}{\bf{G}}_k^{\rm{N,H}}} \right\}} \right]_{m,m}}
		= \sum\limits_{j = 1}^{N^{\rm B}} {\sum\limits_{i = 1}^{N^{\rm R}} {{{\left[ {{{\left| {{{\bf{\Gamma }}_k}} \right|}^2} } \right]}_{i,j}}{\mathbb E}\left\{ {{g_{m,i}}{g_{m,j}^{*}}} \right\}} } \\
		=&  {\sum\nolimits_{i = 1}^{N^{\rm R}} {{{\left[ {{{\left| {{{\bf{\Gamma }}_k}} \right|}^2} } \right]}_{i,i}}{\mathbb E}\left\{ {{g_{m,i}}{g_{m,i}^{*}}} \right\}} } 
		=  \frac{{\left\| {{{\bm{\gamma }}_k}} \right\|^2}\rho_k^2}{\kappa+1}.
	\end{aligned}
\end{equation}
Combining the above analysis, the closed-form expression of $T_2$ can be given by 
\begin{equation}
	{T_2} = T_{23}={{{{\left\| {{{\bm{\gamma }}_k}} \right\|}^2}\rho _k^2}}{\bf{G}}_k^{{\rm{L,H}}}{\bf{G}}_k^{\rm{L}}/({{\kappa  + 1}}).
\end{equation}
\subsection{Closed-form expression for $T_3$}
In this case, ${T_3}$ is unfolded as $ {T_{31}} + {T_{32}} + {T_{33}} + {T_{34}}$, where
\begin{subequations}
 \begin{align}
		{T_{31}}={\mathbb E}\left\{ {\bf{G}}_k^{{\rm{N,H}}}\right\}{\bf{G}}_k^{\rm{L}}{{\left| {{{\bf{\Gamma }}_k}} \right|}^2}{\bf{G}}_k^{{\rm{L,H}}}{\bf{G}}_k^{\rm{L}},\\
		{T_{32}}={\mathbb E}\left\{ {{\bf{G}}_k^{{\rm{N,H}}}{\bf{G}}_k^{\rm{L}}{{\left| {{{\bf{\Gamma }}_k}} \right|}^2}{\bf{G}}_k^{{\rm{L,H}}}{\bf{G}}_k^{\rm{N}}} \right\},\\
		{T_{33}}={\mathbb E}\left\{ {{\bf{G}}_k^{{\rm{N,H}}}{\bf{G}}_k^{\rm{L}}{{\left| {{{\bf{\Gamma }}_k}} \right|}^2}{\bf{G}}_k^{{\rm{N,H}}}} \right\}{\bf{G}}_k^{\rm{L}},\\
		{T_{34}}={\mathbb E}\left\{ {{\bf{G}}_k^{{\rm{N,H}}}{\bf{G}}_k^{\rm{L}}{{\left| {{{\bf{\Gamma }}_k}} \right|}^2}{\bf{G}}_k^{{\rm{N,H}}}{\bf{G}}_k^{\rm{N}}} \right\}.
	\end{align} 
\end{subequations}
Here, $T_{31}=T_{34}={\bf 0}$ holds because $T_{31}$ and $T_{34}$ contain the first- and third-order moments of ${\bf{G}}_k^{\rm{N}}$, respectively. Note that $T_{33}=T_{22}^{\rm H} = {\bf 0}$. Following the analysis procedure for $T_{23}$, we can easily obtain 
\begin{equation}
	{T_3} =T_{32}= {{\rho _k^2}}{\left\| {{\bf{G}}_k^{\rm{L}}{{\bf{\Gamma }}_k}} \right\|^2}{{\bf{I}}_{ {N^{\rm{R}}}}}/({{\kappa  + 1}}).
\end{equation}

\subsection{Closed-form expression for $T_4$}

In the final case, we have ${T_4} = {T_{41}} + {T_{42}} + {T_{43}} + {T_{44}}$, where ${T_{41}}={\mathbb E}\{ {\bf{G}}_k^{{\rm{N,H}}}{\bf{G}}_k^{\rm{N}} \}{{| {{{\bf{\Gamma }}_k}} |}^2}{\bf{G}}_k^{{\rm{L,H}}}{\bf{G}}_k^{\rm{L}}$, ${T_{42}}={\mathbb E}\{ {{\bf{G}}_k^{{\rm{N,H}}}{\bf{G}}_k^{\rm{N}}{{| {{{\bf{\Gamma }}_k}} |}^2}{\bf{G}}_k^{{\rm{L,H}}}{\bf{G}}_k^{\rm{N}}} \}$, and ${T_{43}}={\mathbb E}\{ {{\bf{G}}_k^{{\rm{N,H}}}{\bf{G}}_k^{\rm{N}}{{| {{{\bf{\Gamma }}_k}} |}^2}{\bf{G}}_k^{{\rm{N,H}}}} \}{\bf{G}}_k^{\rm{L}}$ are the conjugated-transpose of $T_{14}$, $T_{34}$, and $T_{24}$, respectively.
The element in the $m$-th row and $c$-th column of $	{T_{44}}={\mathbb E}\{ {{\bf{G}}_k^{{\rm{N,H}}}{\bf{G}}_k^{\rm{N}}{{| {{{\bf{\Gamma }}_k}} |}^2}{\bf{G}}_k^{{\rm{N,H}}}{\bf{G}}_k^{\rm{N}}} \}$ is derived as
\begin{equation}
	\begin{aligned}
		\left[T_{44}\right]_{m,c}&={\sum\limits_{n = 1}^{{N^{\rm{R}}}} {\sum\limits_{t = 1}^{{N^{\rm{R}}}} {\sum\limits_{j = 1}^{{N^{\rm{B}}}} {\sum\limits_{i = 1}^{{N^{\rm{B}}}} {\mathbb E}\left\{{g_{j,m}^{\rm{*}}{g_{j,t}}{[{{\left| {{{\bf{\Gamma }}_k}} \right|}^2}]_{t,n}}g_{i,n}^{\rm{*}}{g_{i,c}}}  \right\}   } } } } \\
		&\mathop  = \limits^{\left( a \right)}  \sum\nolimits_{n = 1}^{{N^{\rm{R}}}} {\sum\nolimits_{j = 1}^{{N^{\rm{B}}}} {\sum\nolimits_{i = 1}^{{N^{\rm{B}}}} {{{[{{\left| {{{\bf{\Gamma }}_k}} \right|}^2}]}_{n,n}}{\mathbb E}\left\{ {g_{j,m}^{\rm{*}}{g_{j,n}}g_{i,n}^{\rm{*}}{g_{i,c}}} \right\}} } } ,
	\end{aligned} 
\end{equation}
where $(a)$ is due to ${{[{{| {{{\bf{\Gamma }}_k}} |}^2}]}_{t,n}}=0$ for $t\ne n$. {When $c\ne m$, $[T_{44}]_{m,c}$ contain the first- and third-order moments of $g_{i,c}$.} Thus, $	[T_{44}]_{m,c}=0$ holds for $c\ne m$. The $m$-th diagonal element of $T_{44}$ can be given by 
\begin{equation}\label{eq-T-44mm}
	\begin{aligned}
		&\left[T_{44}\right]_{m,m}=\sum\nolimits_{j = 1}^{{N^{\rm{B}}}} {\sum\nolimits_{i = 1}^{{N^{\rm{B}}}} {{{[{{\left| {{{\bf{\Gamma }}_k}} \right|}^2}]}_{m,m}}{\mathbb E}\left\{ {g_{j,m}^{\rm{*}}{g_{j,m}}g_{i,m}^{\rm{*}}{g_{i,m}}} \right\}} } \\
		 +& \sum\nolimits_{n \ne m}^{{N^{\rm{R}}}} {\sum\nolimits_{j = 1}^{{N^{\rm{B}}}} {\sum\nolimits_{i = 1}^{{N^{\rm{B}}}} {{{[{{\left| {{{\bf{\Gamma }}_k}} \right|}^2}]}_{n,n}}{\mathbb E}\left\{ {g_{j,m}^{\rm{*}}{g_{i,m}}} \right\}{\mathbb E}\left\{ {{g_{j,n}}g_{i,n}^{\rm{*}}} \right\}} } } .
	\end{aligned} 
\end{equation}
By decoupling $j=i$ and $j\ne i$, \eqref{eq-T-44mm} can be simplified as
\begin{equation}
	\begin{aligned}
		&\left[T_{44}\right]_{m,m}=\sum\limits_{i = 1}^{{N^{\rm{B}}}} {{{[{{\left| {{{\bf{\Gamma }}_k}} \right|}^2}]}_{m,m}}{\mathbb E}\left\{ {\left| {{g_{i,m}}} \right|^4} \right\}} \\
		 + &\sum\limits_{j \ne i}^{{N^{\rm{B}}}} {\sum\limits_{i = 1}^{{N^{\rm{B}}}} {{{[{{\left| {{{\bf{\Gamma }}_k}} \right|}^2}]}_{m,m}}{\mathbb E}\left\{ {g_{j,m}^{\rm{*}}{g_{j,m}}} \right\}{\mathbb E}\left\{ {g_{i,m}^{\rm{*}}{g_{i,m}}} \right\}} } \\
		  + &\sum\limits_{n \ne m}^{{N^{\rm{R}}}} {\sum\limits_{i = 1}^{{N^{\rm{B}}}} {{{[{{\left| {{{\bf{\Gamma }}_k}} \right|}^2}]}_{n,n}}{\mathbb E}\left\{ {g_{i,m}^{\rm{*}}{g_{i,m}}} \right\}{\mathbb E}\left\{ {{g_{i,n}}g_{i,n}^{\rm{*}}} \right\}} } .
	\end{aligned} 
\end{equation}
Given that ${\mathbb E}\{ {| {{g_{i,m}}} |^2} \}=\frac{\rho_k^2}{\kappa+1}$ and ${\mathbb E}\{ {| {{g_{i,m}}} |^4} \}=2\frac{\rho_k^2}{\kappa+1}\frac{\rho_k^2}{\kappa+1}$, the $m$-th diagonal element of $T_{44}$ can be given by
\begin{equation}
	\left[T_{44}\right]_{m,m}=	\left( {{N^{\rm{B}}}{{[{{\left| {{{\bf{\Gamma }}_k}} \right|}^2}]}_{m,m}} + {{\left\| {{{\bm{\gamma }}_k}} \right\|}^2}} \right){{{N^{\rm{B}}}\rho _k^4}}/{{{{\left( {\kappa  + 1} \right)}^2}}}.
\end{equation}
In this case, we can obtain
\begin{equation}
	\begin{aligned}
		&{T_4} = T_{14}^{\rm{H}} + {T_{44}} \\
		&= \frac{{{N^{\rm{B}}}\rho _k^2}}{{\kappa  + 1}}{\left| {{{\bf{\Gamma }}_k}} \right|^2}{\bf{G}}_k^{{\rm{L}},{\rm{H}}}{\bf{G}}_k^{\rm{L}} + \frac{{{N^{\rm{B}}}\rho _k^4}}{{{{(\kappa  + 1)}^2}}}\left( {{N^{\rm{B}}}{{\left| {{{\bf{\Gamma }}_k}} \right|}^2} + {{\left\| {{{\bm{\gamma }}_k}} \right\|}^2}{{\bf{I}}_{ {N^{\rm{R}}}}}} \right)
	\end{aligned}
\end{equation}

By combining closed form results of $T_1$ to $T_4$, the proof of Lemma \ref{Lemma-2} is completed.

\section{}\label{App-fx}
By denoting ${a_1} = {N^{\rm{B}}}\kappa  + 1$, ${a_2} = 2{N^{\rm{B}}}\kappa {\gamma _{\max }}$, and ${a_3} = ( {{N^{\rm{B}}} + {N^{\rm{R}}} + \kappa {N^{\rm{R}}}} )\gamma _{\max }^2-a_1 N^{\rm R}$, \eqref{eq-fx-def} can be rewritten as
\begin{equation}\label{eq-APP-fx}
	f\left( x \right) = a_1+\left({{{a_2}x + {a_3}}}\right)/\left({{\kappa {x^2} + {N^{\rm{R}}}}}\right).
\end{equation}
The derivative of $	f( x ) $ with respect to $x$ is calculated as
\begin{equation}\label{eq-fx'}
	f'\left(x\right)={({ - \kappa {a_2}{x^2} - 2\kappa {a_3}x + {a_2}{N^{\rm{R}}}})}/{{{{\left( {\kappa {x^2} + {N^{\rm{R}}}} \right)}^2}}}.
\end{equation}
For $x  \in [ {0,{N^{\rm{R}}}{\gamma _{\max }}} ]$, we can obtain $	f'\left(0\right)={{{a_2}}}/{{{N^{\rm{R}}}}} > 0$ and 
\begin{equation}
	\begin{aligned}
		&f'\left( {{N^{\rm{R}}}{\gamma _{\max }}} \right) = \\
		&\frac{{2{N^{\rm{R}}}\kappa {\gamma _{\max }}\left( {\frac{{{N^{\rm{B}}}\kappa {N^{\rm{R}}} + {N^{\rm{R}}} + {N^{\rm{B}}}}}{{{N^{\rm{B}}}\kappa {N^{\rm{R}}} + {N^{\rm{R}}} + {N^{\rm{B}}} + \kappa {N^{\rm{R}}}}} - \gamma _{\max }^2} \right)}}{{\left( {{N^{\rm{B}}}\kappa {N^{\rm{R}}} + {N^{\rm{B}}} + {N^{\rm{R}}} + \kappa {N^{\rm{R}}}} \right){{\left( {\kappa {N^{{\rm{R}}{\rm{2}}}}\gamma _{\max }^2 + {N^{\rm{R}}}} \right)}^2}}}\\
		&\mathop  \le \limits^{\left( a \right)} 0,
	\end{aligned}
\end{equation}
where $(a)$ is because $\frac{{{N^{\rm{B}}}\kappa {N^{\rm{R}}} + {N^{\rm{R}}} + {N^{\rm{B}}}}}{{{N^{\rm{B}}}\kappa {N^{\rm{R}}} + {N^{\rm{R}}} + {N^{\rm{B}}} + \kappa {N^{\rm{R}}}}} - \gamma _{\max }^2 < 1 - \gamma _{\max }^2 < 0$. In this case, given that the numerator of \eqref{eq-fx'} is a downward-opening parabolic function, $f'(x)=0$ has a unique solution within $[ {0,{N^{\rm{R}}}{\gamma _{\max }}} ]$. Setting $f'(x)=0$, then the optimal solution $x_{\rm opt}$ that maximizes $f(x)$ can be calculated as 
\begin{equation}
	{x_{{\rm{opt}}}} = {({\sqrt {{\kappa ^2}a_3^2 + \kappa a_2^2{N^{\rm{R}}}}  - \kappa {a_3}})}/{{\kappa {a_2}}}.
\end{equation}
Substituting ${x_{{\rm{opt}}}} $ into \eqref{eq-APP-fx} can obtain \eqref{eq-fxopt}.

\section{ }\label{APP-T2}
{Given $a_1$, $a_2$, and $a_3$ defined in Appendix \ref{App-fx}, applying $\sqrt {{\kappa ^2}a_3^2}  \le \sqrt {{\kappa ^2}a_3^2 + \kappa a_2^2{N^{\rm{R}}}}  \le \sqrt {{\kappa ^2}a_3^2}  + \sqrt {\kappa a_2^2{N^{\rm{R}}}} $ to $f(x_{\rm opt})$ in \eqref{eq-fxopt} yields}
\begin{equation}\label{eq-APP-T2-fxopt}
	\begin{aligned}
		f\left( {{x_{{\rm{opt}}}}} \right) 	&< {a_1} + \frac{{a_2^2 \left(\sqrt {{\kappa ^2}a_3^2}  + \sqrt {\kappa a_2^2{N^{\rm{R}}}} \right) }}{{{{\left( {\sqrt {{\kappa ^2}a_3^2}  - \kappa {a_3}} \right)}^2} + {N^{\rm{R}}}\kappa a_2^2}}\\
		&= {a_1} + \left({{\left| {{a_3}} \right| + 2{N^{\rm{B}}}{\gamma _{\max }}\sqrt {\kappa {N^{\rm{R}}}} }}\right)/{{{N^{\rm{R}}}}}.
	\end{aligned}
\end{equation}
For a sufficiently large reflection amplitude satisfying
\begin{equation}
	 \gamma _{\max }^2 > {{\left( {{N^{\rm{B}}}\kappa  + 1} \right){N^{\rm{R}}}}}/\left({{{N^{\rm{B}}} + {N^{\rm{R}}} + \kappa {N^{\rm{R}}}}}\right),
\end{equation} we have $| {{a_3}} |=a_3$ and \eqref{eq-APP-T2-fxopt} can be expresses as
\begin{equation}\label{eq-APP-T2-f1}
		f\left( {{x_{{\rm{opt}}}}} \right) <\frac{{\left( {{N^{\rm{B}}} + {N^{\rm{R}}} + \kappa {N^{\rm{R}}}} \right)\gamma _{\max }^2 + 2{N^{\rm{B}}}{\gamma _{\max }}\sqrt {\kappa {N^{\rm{R}}}} }}{{{N^{\rm{R}}}}}.
\end{equation}
{When $\gamma _{\max }^2 \le{{\left( {{N^{\rm{B}}}\kappa  + 1} \right){N^{\rm{R}}}}}/{({{N^{\rm{B}}} + {N^{\rm{R}}} + \kappa {N^{\rm{R}}}})}$, we have $| {{a_3}} |=-a_3$. In this small reflection amplitude regime, $-{a_3} =a_1 N^{\rm R} -( {{N^{\rm{B}}} + {N^{\rm{R}}} + \kappa {N^{\rm{R}}}} )\gamma _{\max }^2<a_1 N^{\rm R}$ is applied to relax the upper bound for more insights. Thus, \eqref{eq-APP-T2-fxopt} can be further bounded by
\begin{equation}\label{eq-APP-T2-f2}
		f\left( {{x_{{\rm{opt}}}}} \right) <\frac{{2\left( {{N^{\rm{B}}}\kappa  + 1} \right){N^{\rm{R}}} + 2{N^{\rm{B}}}{\gamma _{\max }}\sqrt {\kappa {N^{\rm{R}}}} }}{{{N^{\rm{R}}}}}.
\end{equation}
Note that the use of $-{a_3} <a_1 N^{\rm R}$ makes \eqref{eq-APP-T2-f2} looser than \eqref{eq-APP-T2-f1}, causing a discontinuity for ${\beta _{k,i}^{{\rm Upp},2}}$ in $\gamma _{\max }^2 ={{\left( {{N^{\rm{B}}}\kappa  + 1} \right){N^{\rm{R}}}}}/{({{N^{\rm{B}}} + {N^{\rm{R}}} + \kappa {N^{\rm{R}}}})}$.} On the basis of \eqref{eq-APP-T2-f1}, \eqref{eq-APP-T2-f2}, we have ${\beta _{k,i}^{{\rm Upp},1}}<{\beta _{k,i}^{{\rm Upp},2}}$.

\section{Proof of Lemma \ref{L-Rec}}\label{APP-L-Rec}
By unfolding ${{\bf{h}}_i} = {[ {{\bf{h}}_{1,i}^{\rm{H}},{\bf{h}}_{2,i}^{\rm{H}},{\bf{h}}_{3,i}^{\rm{H}}} ]^{\rm{H}}}$ and ${{\bf{h}}_{k,i}} = {{\bf{G}}_k}{{\bf{\Gamma }}_k}{{\bf{g}}_{k,i}}$, the average received signal power can be expressed as
\begin{equation}
	\begin{aligned}
		&{\mathbb E}\left\{ {{{\left\| {{{\bf{h}}_i}} \right\|}^4}} \right\} \\
		=& \sum\limits_{k = 1}^3 {\sum\limits_{m = 1}^3 {\mathbb E}{\left\{ {{\bf{g}}_{k,i}^{\rm{H}}{\bf{\Gamma }}_k^{\rm{H}}{\bf{G}}_k^{\rm{H}}{{\bf{G}}_k}{{\bf{\Gamma }}_k}{{\bf{g}}_{k,i}}{\bf{g}}_{m,i}^{\rm{H}}{\bf{\Gamma }}_m^{\rm{H}}{\bf{G}}_m^{\rm{H}}{{\bf{G}}_m}{{\bf{\Gamma }}_m}{{\bf{g}}_{m,i}}} \right\}} } .
	\end{aligned}
\end{equation}

When $m\ne k$, we rewrite ${\mathbb E}\{ {{{\left\| {{{\bf{h}}_i}} \right\|}^4}} \}$ as
\begin{equation}\label{APP-L-Rec-mk}
	\begin{aligned}
		{\mathbb E}\left\{ {{{\left\| {{{\bf{h}}_i}} \right\|}^4}} \right\}=&\sum\nolimits_{k = 1}^3 {\bf{g}}_{k,i}^{\rm{H}}{\bf{\Gamma }}_k^{\rm{H}}\left\{ {{\bf{G}}_k^{\rm{H}}{{\bf{G}}_k}} \right\}{{\bf{\Gamma }}_k}{{\bf{g}}_{k,i}}\\
			&\times\sum\nolimits_{m \ne k}^3 {{\bf{g}}_{m,i}^{\rm{H}}{\bf{\Gamma }}_m^{\rm{H}}\left\{ {{\bf{G}}_m^{\rm{H}}{{\bf{G}}_m}} \right\}{{\bf{\Gamma }}_m}{{\bf{g}}_{m,i}}}.  
	\end{aligned}
\end{equation}
Utilizing results in Lemma \ref{Lemma-1}, \eqref{APP-L-Rec-mk} can be derived as
\begin{equation}\label{APP-L-Rec-mk-r}
	{\mathbb E}\left\{ {{{\left\| {{{\bf{h}}_i}} \right\|}^4}} \right\} = \frac{{{N^{{\rm{B,2}}}}}}{{{{\left( {\kappa  + 1} \right)}^2}}}\sum\nolimits_{k = 1}^3 {\rho _{k,i}^2\rho _k^2{F_{0,k,i}}} \sum\nolimits_{m \ne k}^3 {\rho _{m,i}^2\rho _m^2{F_{0,m,i}}}  .
\end{equation}

When $m=k$, ${\mathbb E}\{ {{{\left\| {{{\bf{h}}_i}} \right\|}^4}} \}$ can be simplified as
\begin{equation}\label{APP-L-Rec-kk}
	\begin{aligned}
		&{\mathbb E}\left\{ {{{\left\| {{{\bf{h}}_i}} \right\|}^4}} \right\} \\
		=&  {\sum\nolimits_{k = 1}^3{\bf{g}}_{k,i}^{\rm{H}}{\bf{\Gamma }}_k^{\rm{H}} {\mathbb E}{\left\{ {\bf{G}}_k^{\rm{H}}{{\bf{G}}_k}{{\bf{\Gamma }}_k}{{\bf{g}}_{k,i}}{\bf{g}}_{m,i}^{\rm{H}}{\bf{\Gamma }}_k^{\rm{H}}{\bf{G}}_k^{\rm{H}}{{\bf{G}}_k} \right\}{{\bf{\Gamma }}_k}{{\bf{g}}_{k,i}}} } ,
	\end{aligned}
\end{equation}
which has a similar expression  to ${\mathbb E}\{ {{{| {{{\bf{w}}^{\rm{H}}}{{\bf{z}}^{\rm{R}}}} |}^2}} \}$ in \eqref{eq-EwHzR}. By replacing ${{\bf{\Gamma }}_k}$ in \eqref{eq-EwHzR} with ${{\bf{\Gamma }}_k}{{\bf{g}}_{k,i}}$ and following the proof procedure of Theorem \ref{Theo-1}, \eqref{APP-L-Rec-kk} can  readily derived as follows:
\begin{equation}\label{APP-L-Rec-kk-r}
	\begin{aligned}
		&{\mathbb E}\left\{ {{{\left\| {{{\bf{h}}_i}} \right\|}^4}} \right\}\\
		=&\frac{{{N^{\rm{B}}}}}{{{{\left( {\kappa  + 1} \right)}^2}}}\sum\limits_{k = 1}^3 {\rho _{k,i}^4\rho _k^4\left( {\left( {{N^{\rm{B}}} + 1} \right)F_{0,k,i}^2 - {\kappa^2}{{\left| {{\bm{\gamma }}_k^{\rm{H}}{\bf{a}}_{k,i}^{\rm{R}}} \right|}^4}} \right)} .
	\end{aligned}
\end{equation}
Merging \eqref{APP-L-Rec-mk-r} and \eqref{APP-L-Rec-kk-r}, we complete the proof.

\section{ }\label{APP-last}
{With the closed-form expressions of the optimized ${\bm{\omega}}_k$ and ${\bm{\alpha}}_k$ available, substituting the identical BS-RIS channel path loss $\rho_0=\rho_k$ ($\forall k$) into $\eta_{\rm active}$ yields
\begin{equation}
	\begin{aligned}
	\eta_{\rm active} &= \frac{{{E_i}{\mathbb{E}}\left\{ {{{\left\| {{{\bf{h}}_i}} \right\|}^4}} \right\}}}{{{\mathbb{E}}\left\{ {{{\left| {{{\bf{w}}^{\rm{H}}}{{\bf{z}}^{\rm{R}}}} \right|}^2}} \right\} + \sigma _{\rm{B}}^2{\mathbb{E}}\left\{ {{{\left\| {{{\bf{h}}_i}} \right\|}^2}} \right\}}}\\
	 &= \frac{{{C_1}\gamma _{\max }^2}}{{{C_2}\frac{{\sigma _{\rm{R}}^2\gamma _{\max }^2}}{{{d^2}}}\sum\nolimits_{k = 1}^3 {\rho _{k,i}^2}  + {C_3}\sum\nolimits_{k = 1}^3 {\rho _{k,i}^2} }},
		\end{aligned}
	\end{equation}
where 
\begin{equation}
	\begin{aligned}
		C_1 =&  \frac{{\rho _0^2{E_i}{N^{\rm{B}}}{N^{\rm{R}}}{{\left( {\kappa {N^{\rm{R}}} + 1} \right)}^2}}}{{\kappa  + 1}}\sum\nolimits_{k = 1}^3 {\sum\nolimits_{m = 1}^3 {\rho _{k,i}^2\rho _{m,i}^2} } \\
		 &+ \frac{{2\kappa {N^{\rm{R}}} + 1}}{{\kappa  + 1}}\sum\nolimits_{k = 1}^3 {\rho _{k,i}^4}, \\
		 C_2  =& \left( {{N^{\rm{B}}}\left( {{\kappa ^2}{N^{{\rm{R}},2}} + 2\kappa {N^{\rm{R}}} + 1} \right) + {N^{\rm{R}}}\left( {\kappa {N^{\rm{R}}} + \kappa  + 1} \right)} \right)\\
		 	&\times \frac{{{\lambda ^2}}}{{16{\pi ^2}\left( {\kappa  + 1} \right)}},\\
		 	C_3 = & {\sigma _{\rm{B}}^2\left( {\kappa {N^{\rm{R}}} + 1} \right)}.
	\end{aligned}
	\end{equation}
By setting $\sigma _{\rm{R}}^2 = 0$ and ${\gamma _{\max }} = 1$, we have 
\begin{equation}
		{\eta_{\rm passive}} = \frac{{{C_1}}}{{ {C_3}\sum\nolimits_{k = 1}^3 {\rho _{k,i}^2} }}.
\end{equation}
When $\eta_{\rm active}<\eta_{\rm passive}$, \eqref{eq:d_min} can be obtained.
}

%\section*{Acknowledgment}

\ifCLASSOPTIONcaptionsoff
  \newpage
\fi

%\end{IEEEbiography}


%\newpage


\end{document}